\documentclass[11pt]{article}
\pdfoutput=1
\usepackage[margin=1.0in]{geometry}
\usepackage{amsmath,amssymb,graphicx}
\usepackage{hyperref}
\usepackage{slashed}
\usepackage{bm}

\usepackage[T1]{fontenc}

\usepackage[utf8]{inputenc}

\usepackage{color,soul}

\usepackage[greek,english]{babel}
\newcommand{\cpi}{\text{\greektext p}}
\newcommand{\cepsilon}{\text{\greektext e}}

\usepackage[font=small,labelfont=bf]{caption}
\usepackage{cite}

\usepackage{tabu}

\newcommand{\iu}{{\mathrm i}}
\newcommand{\E}{{\mathrm e}}

\newcommand{\tF}{{\widetilde F}}
\newcommand{\tG}{{\widetilde G}}
\newcommand{\tQ}{{\widetilde Q}}
\newcommand{\tL}{{\widetilde L}}

\newcommand{\cL}{{\cal L}}

\newcommand{\bbZ}{\mathbb{Z}}

\newcommand{\df}{\mathrel{:=}}

\newcommand{\Tr}{\text{Tr}}
\newcommand{\p}{\partial}
\newcommand{\Lag}{\mathcal{L}}

\newcommand{\be}{\begin{equation}}
\newcommand{\ee}{\end{equation}}

\usepackage{xcolor}

\title{\bf Axion Periodicity and Coupling Quantization in the Presence of Mixing}
\author{Katherine Fraser and Matthew Reece\\
{\small \color{gray} \texttt{kfraser, mreece~(@g.harvard.edu)}}\\
{\small Department of Physics, Harvard University, Cambridge, MA, 02138}}

\begin{document}
\maketitle

\begin{abstract}
Mixing of axion fields is widely used to generate EFTs with phenomenologically advantageous features, such as hierarchies between axion couplings to different gauge fields and/or large effective field ranges. While these features are strongly constrained by periodicity for models with only a single axion, mixing has been used in the literature (sometimes incorrectly) to try to evade some of these constraints. In this paper, we ask whether it is possible to use axion mixing to generate an EFT of axions that evades these constraints by flowing to a theory of a non-compact scalar in the IR. We conclude that as long as the light axion is exactly massless, it will inherit the periodicity and associated constraints of the UV theory. However, by giving the light axion a mass, we can relax these constraints with effects proportional to the axion mass squared, including non-quantized couplings and the realignment of monodromy to a light axion with a larger field range. To show this, we consider various examples of axions mixing with other axions or with non-compact scalar fields, and work in a basis where coupling quantization is manifest. This basis makes it clear that in the case where an axion is eaten through the Higgs or St{\"u}ckelberg mechanism, the light axion does not have a large effective field range, in contrast to some recent claims in the literature. Additionally, we relate our results about axion EFTs to a well-known fact about gauge theory: that QFTs with compact gauge groups in the UV flow to QFTs with compact gauge groups in the IR, and make this correspondence precise in the 2+1 dimensional case.
\end{abstract}

\section{Introduction}

Axion fields are ubiquitous in theories of physics beyond the Standard Model. For our purposes, the defining feature of an axion (as compared with a generic scalar field) is that it is a compact boson, whose target space is a circle. That is, an axion field by definition is identified under a discrete shift symmetry:
\be
a(x) \cong a(x) + 2 \cpi n F_a,  \quad n \in \bbZ   \label{eq:ashift}
\ee
where $2\cpi F_a$ is the fundamental period of the field. The canonical example is the QCD axion, which provides a dynamical explanation of the lack of observed CP violation in the strong interactions \cite{Peccei:1977ur, Peccei:1977hh, Weinberg:1977ma, Wilczek:1977pj}; for more up-to-date reviews, see \cite{Kim:2008hd, Graham:2015ouw, Hook:2018dlk}. Axions have also been postulated to: play the role of the inflaton that drove exponential expansion in the early Universe \cite{Freese:1990rb}; account for the presence of large-scale magnetic fields in intergalactic space \cite{Turner:1987bw, Garretson:1992vt}; serve as dark matter \cite{Preskill:1982cy, Dine:1982ah, Abbott:1982af}; or help generate the matter/antimatter asymmetry in the Universe \cite{Alexander:2004us}. The wide range of roles that axions can play in cosmology is reviewed in \cite{Marsh:2015xka}. The existence of axion fields, possibly in large numbers, is a prediction of string theory \cite{Witten:1984dg, Barr:1985hk, Choi:1985je, Dine:1986bg, Svrcek:2006yi}, and in recent years there has been significant interest in the phenomenology of theories with many axions \cite{Arvanitaki:2009fg}. 

The periodicity \eqref{eq:ashift} of an axion has significant consequences for the structure of axion effective field theories. An immediate consequence is that axion potentials must be periodic functions. Less obviously, for reasons that we will review below, couplings of the form $a F_{\mu \nu} \tF^{\mu \nu}$ of axions to gauge fields must have {\em quantized} coefficients, which are integer multiples of $e^2/(16\cpi^2 F_a)$. This poses a significant challenge for many phenomenological models that rely on axions. For example, in a cosmological model one might be interested in an axion potential with a very large field range, but at the same time may want a large coupling of the axion to gauge fields (e.g., for reheating \cite{Adshead:2015pva}, magnetogenesis \cite{Turner:1987bw, Garretson:1992vt}, or for the structure of the inflationary model itself \cite{Anber:2009ua,Adshead:2012kp}).  Because the axion potential and the $aF\tF$ couplings both depend on $F_a$, our options for building such models are very limited, unless the constraints imposed by periodicity \eqref{eq:ashift} can be relaxed.

In this paper, our chief interest is in the robustness of the constraints associated with axion periodicity. Can an effective field theory containing periodic axions flow in the infrared to a new effective field theory in which some of the axion fields have become effectively non-compact, and hence have fewer constraints on their couplings? By considering various examples, in which axions mix with other axions or with non-compact scalar fields, we will argue that the options are very limited. In particular, we claim that whenever some of the axions remain massless in the IR, they will continue to {\em exactly} respect  periodicity constraints. Deviations from these constraints are always proportional to powers of the axion mass. This is reminiscent of the fact that quantum field theories with compact gauge groups in the UV flow to quantum field theories with (possibly different) compact gauge groups in the IR. As we will discuss below, this is more than a superficial similarity.

Before summarizing our results in more detail, let us briefly review the properties of axion effective field theories enforced by the shift symmetry \eqref{eq:ashift}.

\subsection{Review: quantized couplings in axion EFT}

Readers who are thoroughly familiar with the reason why $aF\tF$ couplings are quantized, and how to precisely formulate this condition in theories with fermions, can safely skip this subsection, though it may be useful for establishing our conventions.

Because we will be studying scenarios in which axions may not have canonical kinetic terms, it is often useful to consider {\em dimensionless} axion fields $\theta$ which are normalized to have period $2\cpi$,
\be
\theta(x) \cong \theta(x) + 2 \cpi n,  \quad n \in \bbZ.  \label{eq:thetashift}
\ee
These identifications on field space may be thought of as discrete gauge symmetries. In certain theories, such gauge symmetries may be {\em spontaneously broken}, in which case an axion may appear to acquire a non-periodic potential or other interactions that violate the symmetry. In such cases, there is a monodromy when the axion traverses its fundamental circle, so that the full set of states of the theory actually respects the underlying symmetry. We will refer to such fields as ``monodromy axions.'' Monodromy axions have played a major role in inflationary model-building \cite{Silverstein:2008sg, Mcallister:2008hb}.

The periodicity \eqref{eq:thetashift} imposes important, well-known constraints on the effective field theory of an axion. An obvious one is that (in the absence of monodromy) the potential is periodic, $V(\theta) = V(\theta + 2\cpi)$. In many theories of axions, there are important couplings between axions and gauge fields of the form
\be
\cL_{\theta F \tF} = k\frac{\theta}{16\cpi^2} F_{\mu \nu} \tF^{\mu \nu}, \quad k \in \bbZ, \label{eq:LthetaFF}
\ee
where the dual gauge field is defined as $\tF^{\mu \nu} \df \frac{1}{2} \cepsilon^{\mu\nu\rho\sigma} F_{\rho \sigma}$. The requirement that $k$ is quantized follows from the axion periodicity \eqref{eq:thetashift}. Here we have assumed that the normalization of $F$ is such that the kinetic term is $-\frac{1}{4e^2} F_{\mu \nu}F^{\mu \nu}$ and that particles of gauge charge $q \in \bbZ$ couple to the gauge field through the action $S = q \int_\gamma A$, where $\gamma$ is the charged-particle worldline and $A = A_\mu\, {\rm d}x^\mu$ is the 1-form gauge field.
The reason that the coupling $k$ in \eqref{eq:LthetaFF} is quantized is that the interaction Lagrangian is not gauge invariant: its coefficient changes value under the shift $\theta \mapsto \theta + 2\cpi n$. However, the path-integral measure is well-defined whenever $k \in \bbZ$, because $\exp(\iu \int {\rm d}^4x\,\cL_{\theta F \tF})$ is well-defined.

Our statement of the quantization of the coupling \eqref{eq:LthetaFF} applies when we consider this coupling in isolation. In theories with fermions that couple to $\theta$ and transform under the gauge field, the correct statement of coupling quantization refers to an invariant combination of couplings. For example, if we consider a Lagrangian containing the terms
\be
\iu \overline{\Psi} \slashed{D} \Psi + c_\partial (\partial_\mu \theta) \overline{\Psi} \gamma^\mu \gamma^5 \Psi - \left[m \E^{\iu c_m \theta} \overline{\Psi}_L \Psi_R + \text{h.c.}\right] + c_F \frac{\theta}{16\cpi^2} F_{\mu \nu} \tF^{\mu \nu},   \label{eq:thetaFFfermions}
\ee
where the $\Psi$ transform in the fundamental representation of the gauge group, then the field redefinition
\be
\Psi_L \mapsto \E^{\iu a \theta} \Psi_L, \quad \Psi_R \mapsto \E^{-\iu a \theta} \Psi_R
\ee
produces a different Lagrangian with replacements
\be
c_\partial \mapsto c_\partial - a, \quad c_m \mapsto c_m - 2 a, \quad c_F \mapsto c_F + 2 a, \label{eq:fieldredefcoupling}
\ee
with the shift in $c_F$ arising due to the chiral anomaly (e.g., from the anomalous transformation of the fermion measure in the path integral). As a result, it is clearly not correct to demand that $c_F \in \bbZ$ in general. However, if we first decouple the axion from the fermions by performing a field redefinition to set $c_m = 0$, so that the axion couples {\em only} through interactions like $c_\partial$ that preserve a continuous shift symmetry and through $\theta F \tF$ type terms, {\em then} the latter terms are quantized. In other words, the correct quantization condition in the case of the Lagrangian \eqref{eq:thetaFFfermions} is
\be
c_m + c_F \in \bbZ.   \label{eq:invariantquantization}
\ee
This suffices to ensure that the path integral is well-defined under the identification \eqref{eq:thetashift}. Invariant combinations of couplings including the derivative term, such as $c_\partial - \frac{1}{2} c_m$, can take any real value.

The quantization rules \eqref{eq:LthetaFF} or \eqref{eq:invariantquantization} apply for axion couplings to U(1) gauge fields or to nonabelian gauge fields, up to a change in the linear combination of coefficients appearing in \eqref{eq:invariantquantization} that depends on the Dynkin index of the gauge representation of the fermions. In most of the equations in our paper, a factor of $F_{\mu \nu} \tF^{\mu \nu}$ may be replaced by $\frac{1}{2} F^a_{\mu \nu} \tF^{a\mu \nu}$ for a nonabelian group without changing the validity of our statements. The only necessarily abelian gauge fields that we discuss will be those in \S\ref{sec:spin1mix} that eat axions to acquire a mass, and the higher-dimensional gauge field in \S\ref{sec:non-compact} that is used to engineer a simple scenario with monodromy. (In both cases one could consider nonabelian extensions, but this would complicate the physics without obvious dividends.)

\subsection{Summary: motivation and results}

Axions, like more general scalar fields, can mix with other fields in a variety of ways. They may have mass or kinetic mixing with other axions (e.g., \cite{Babu:1994id,Kim:2004rp,Bachlechner:2014hsa}). Some linear combinations of the axions may be eaten by massive spin-1 fields (via the Higgs mechanism or St{\"u}ckelberg couplings, e.g., \cite{Cheng:2001ys, Anastasopoulos:2006cz, Shiu:2015uva,Shiu:2015xda,Choi:2019ahy}). Axions may even mix with other fields that are not periodic, whether these are ordinary scalar fields or monodromy axions (e.g., \cite{Berg:2009tg}). When some of these fields acquire mass, we can integrate them out to obtain an effective field theory involving only the light fields.

The central question of this paper is: does the EFT of the light fields always inherit a periodicity condition like \eqref{eq:thetashift} and the associated constraints? For example, can one begin with a theory of two axions, one linear combination of which acquires a mass (either through a potential or through being eaten by a massive spin-1 field) so that the remaining, light combination is no longer an axion (i.e., has no well-defined period)? The answer to an analogous question in gauge theory is familiar: if we consider a theory with a compact gauge group, which is reduced to a smaller gauge group in the infrared through Higgsing, then the infrared gauge group will still be compact. For example, in the Standard Model, the photon couples to an electromagnetic charge whose quantization is inherited from the quantization of SU(2)$_L$ and U(1)$_Y$ charges. This follows from the fact that the Higgs field itself carries quantized charges. Similarly, even in theories with kinetic mixing, there is a discrete charge lattice for the {\em massless} U(1) bosons, whether or not they mix with massive spin-1 bosons \cite{Holdom:1985ag, Shiu:2013wxa}. Despite the existence of such analogous results, we emphasize that our results for spin-0 bosons do {\em not} all precisely map to familiar results for spin-1 bosons. For example, we will discuss cases in which spin-0 bosons are eaten by spin-1 bosons, quantization of $a F \tF$ couplings and the role of massless chiral fermions in determining the invariant quantized couplings, and axion monodromy. These additional ingredients require different arguments from those of \cite{Holdom:1985ag, Shiu:2013wxa}.

Apart from its intrinsic interest as a question about the structure of quantum field theory, our motivation for studying this question is that the constraints imposed by the periodicity \eqref{eq:thetashift} can provide serious obstructions to building interesting phenomenological models. The literature on applications of axions in phenomenology is vast, so we cannot provide a complete bibliography, but some of the main themes and specific examples to which our work is relevant include:
\begin{itemize}
\item {\bf Hierarchies between couplings.} One interesting goal is to have axion couplings to $F\tF$ terms with very different sizes. In a theory where these couplings are quantized, this can only be achieved by invoking a large integer, which one could then attempt to explain from within a UV completion (e.g., \cite{Kim:2004rp,Choi:2014rja, Choi:2015fiu, Kaplan:2015fuy}). An obvious application is to the QCD axion, where one might like to separate the coupling to gluons (which determines the axion mass) from the coupling to photons (which is often invoked to provide experimental tests of the theory). Various models can alter the ratio of these couplings \cite{Farina:2016tgd, Agrawal:2017cmd}.
\item {\bf Achieving large field ranges.} Especially in cosmological applications, it is often of great interest to have a field that can evolve over a long distance in field space. For example, this is necessary to produce large primordial gravitational wave signals from standard inflation models \cite{Lyth:1996im}, or to allow novel mechanisms like dynamical relaxation of the weak scale to operate \cite{Graham:2015cka}. In the context of string theory, it is known to be difficult to find axions with fundamental period larger than the Planck scale (e.g., \cite{Banks:2003sx, ArkaniHamed:2006dz, Ooguri:2006in, Rudelius:2014wla}), which has motivated many efforts to build models where small field ranges in the UV become large field ranges in the IR (which are too numerous to review here).  
\item {\bf Reconciling a large field range with a large coupling.} In some cases, the challenge is a blend of the two previous ideas. One might want a large axion field range $f$ appearing in terms like $\cos(a/f)$, but also a large coupling $\frac{\alpha}{8\cpi} \frac{a}{f'} F \tF$, and hence a small scale $f'$. Because $f$ and $f'$ are both related to the axion period, again, it can be difficult to achieve a large separation of these scales. This issue arises in chromonatural inflation \cite{Adshead:2012kp}, which in any single-axion model requires an enormous integer to appear in the effective action \cite{Heidenreich:2017sim, Agrawal:2018mkd}. Similar issues arise when using axion couplings to gauge fields for preheating \cite{Adshead:2015pva}, to suppress the axion dark matter abundance \cite{Agrawal:2017eqm, Kitajima:2017peg}, or to produce dark photon dark matter \cite{Agrawal:2018vin, Co:2018lka, Bastero-Gil:2018uel}. 
\end{itemize}
Separate from these specific phenomenological or model-building goals, if an axion field is discovered experimentally in the future, precisely measuring its couplings and understanding whether they are quantized could play a critical role in interpreting the signal. Clearly, it is important to understand our theoretical expectations before any such discoveries are made.

We will see that in studying simple theories in which multiple axions mix, interesting subtleties arise in examining the periodicity and couplings of a light axion. If one simply examines formulas that are present in the literature, one might suspect that the IR theory in general does not inherit any periodicity constraint from the UV theory. We will encounter a case in which the light axion field appears to be non-compact, and yet inherits periodic couplings just as a compact field would. We will also encounter a case in which the light axion field at first sight appears to be compact, but periodicity-violating couplings appear in the EFT. These results provide tantalizing hints for the construction of phenomenological models that can evade various constraints, and in some cases claims of large effective field ranges in such models have been made in the literature \cite{Shiu:2015uva, Shiu:2015xda,Fonseca:2019aux}. However, in every case that we study, a more careful examination reveals that the periodicity of the axion field and quantization of (properly defined, invariant) couplings are properties of the infrared theory whenever the light axions remain massless. In the particular cases referenced above in \cite{Shiu:2015uva, Shiu:2015xda,Fonseca:2019aux}, the authors overlook subtleties related to the absence of anomalies, which relates the various parameters in the Lagrangian and enforces quantization. In particular, these relations prevent some of the scenarios discussed in \cite{Shiu:2015uva, Shiu:2015xda,Fonseca:2019aux} from being able to generate large effective field ranges. Once a mass is generated, the constraints are loosened. However, all such effects are proportional to powers of the light axion mass.

In order to achieve hierarchies between an axion's coupling to different gauge fields, or between an axion field range and the scale suppressing its coupling to a gauge field, we find the following options:
\begin{itemize}
\item The axion couplings remain quantized due to periodicity, and the hierarchy arises from a large integer, as in the clockwork scenario \cite{Kim:2004rp,Dvali:2007hz, Choi:2014rja, Hebecker:2015rya, Choi:2015fiu, Kaplan:2015fuy}.
\item The axion is massive, and its couplings deviate significantly from their expected quantization due to mixing with other axions with masses generated at the same scale. This possibility is familiar from the QCD axion's coupling to the photon, which obtains a non-quantized contribution from mixing with the $\pi^0$ \cite{Kaplan:1985dv, Srednicki:1985xd, Georgi:1986df}. 
\item Mixing between monodromy axions and ordinary axions can ``realign'' monodromy to a light axion with a larger field range than the original monodromy axion, as in the ``Dante's Inferno'' model \cite{Berg:2009tg}. This effectively extends the axion field range by allowing it to ``unwind.'' 
\end{itemize}
Some aspects of our claims have been noted in other recent work, including \cite{Agrawal:2017cmd} by one of us and \cite{Choi:2019ahy}. We extend earlier work by surveying a wider range of models, but also by situating the question in the broader theoretical context of compactness of gauge groups. Some of our arguments in \S\ref{sec:cosinemix} resemble those made in the past about mixing of spin-1 gauge fields \cite{Holdom:1985ag, Shiu:2013wxa}, though various details (e.g., our use of the Smith normal form in \S\ref{subsec:Naxiongen}, or the effects of turning on a mass for the light axion) are not directly analogous to results in those references.

The outline of this paper is as follows: in \S\ref{sec:cosinemix}, we discuss scenarios in which some linear combinations of axions obtain periodic potentials. We show that the remaining, light scalar fields are always periodic (their field space is a torus) and their couplings are quantized as expected. In \S\ref{sec:spin1mix}, we consider the possibility that a linear combination of axions decouples because it is eaten by a massive spin-1 field (either via the Higgs or St{\"u}ckelberg mechanisms). Again, we show that the uneaten combination is a periodic field with quantized couplings. The results of this section were also obtained independently in \cite{Choi:2019ahy}, which appeared while this paper was being completed. In \S\ref{sec:non-compact}, we discuss the possible mixing of axions with other, non-compact scalars. We show that a theory in which a monodromy axion mixes with a heavier ordinary axion can lead to a ``realignment'' of monodromy to a linear combination of the original axions, so that the axion decay constant is larger in the low-energy effective field theory. In \S\ref{sec:gaugetheorycompare}, we discuss the relationship between our studies of compactness in axion field spaces and the question of compactness of gauge groups. In particular, we point out that in some cases these questions are related by Hodge (electric/magnetic) duality. We suggest that our results fit into a larger picture in which theories with compact gauge groups in the UV always flow to theories with compact gauge groups in the IR. Finally, we very briefly conclude in \S\ref{sec:conclusions}.

\section{Mixing with a Heavier Axion with a Periodic Potential}
\label{sec:cosinemix}

\subsection{Light axion remaining massless}

The first scenario we will consider is when two axions mix and a periodic potential gives a mass to one linear combination of them, leaving one massless axion in the IR. We will argue that there is a consistent EFT description in which the light axion is periodic and has quantized couplings to gauge fields. Elements of our discussion, involving the diagonalization of kinetic mixing in the case of a massive axion, have previously appeared in \cite{Babu:1994id,Higaki:2014qua}, and some of the conclusions about quantized couplings were previously emphasized in \cite{Agrawal:2017cmd}. Nonetheless, it is useful to highlight a confusing aspect of the calculation that has not previously been emphasized, and then explain how this confusion is resolved. We will encounter a similarly confusing intermediate result in \S\ref{sec:spin1mix}, which our experience in this section will help to resolve correctly.

\subsubsection{Setting up the problem in a convenient lattice basis}
\label{subsec:latticebasis}

We will denote our two axion fields $\theta_1$ and $\theta_2$ and assume that they both have period $2\cpi$. A different way to say this is that our field space is a torus, obtained by taking the quotient of the plane $(\theta_1, \theta_2)$ by the lattice $(2\cpi n_1, 2\cpi n_2),$ $n_i \in \bbZ$. A linear transformation
\be
\begin{pmatrix} \theta'_1 \\ \theta'_2 \end{pmatrix} = \begin{pmatrix} a & b \\ c & d \end{pmatrix} \begin{pmatrix} \theta_1 \\ \theta_2 \end{pmatrix}
\ee
preserves this structure provided that 
\be
\begin{pmatrix} a & b \\ c & d \end{pmatrix}  \in \text{GL}(2,\bbZ).
\ee
We will call any such basis for our field space a ``lattice basis.'' Other bases are, of course, possible, but they require us to reparametrize the lattice of identifications of the plane.

We will consider an effective Lagrangian of the form
\be
{\cal L} = -\frac{1}{4e^2} F_{\mu \nu}F^{\mu \nu} + K_{ij} \partial_\mu \theta_i \partial^\mu \theta_j - V(j_1 \theta_1 + j_2 \theta_2) + \frac{k_1 \theta_1 + k_2 \theta_2}{16\cpi^2} F_{\mu \nu} \tF^{\mu \nu},
   \label{eq:Lmassmix}
\ee
where $j_i, k_i \in \bbZ$ but $K_{ij}$ is an arbitrary symmetric real matrix of rank 2. For concreteness, one could imagine the potential to take the form
\be
V(j_1 \theta_1 + j_2 \theta_2) = \Lambda^4 \left[1 - \cos(j_1 \theta_1 + j_2 \theta_2)\right],
\ee
which might be the leading approximation to the potential generated by a confining, pure glue sector via the coupling
\be
\frac{j_1 \theta_1 + j_2 \theta_2}{32\cpi^2} G^a_{\mu \nu} \tG^{a\mu\nu}.
\ee
However, the only important assumption we will make is that $V(x)$ has a minimum at $x = 0$, and a Taylor expansion $V(x) \approx V_0 + \frac{1}{2} \mu^4 x^2 + {\cal O}(x^3)$. Without loss of generality, we will assume that $\gcd(j_1, j_2) = 1$, by absorbing any common factor into the normalization of the function $V$.

In the subsequent discussion, we will often drop the $-\frac{1}{4e^2} F_{\mu \nu}F^{\mu \nu}$ term when writing our Lagrangian. It is understood to be present whenever a coupling to $F\tF$ appears.

It is always possible to perform a GL$(2,\bbZ)$ transformation so that the massive axion is a basis vector, $\theta'_1 = j_1 \theta_1 + j_2 \theta_2$. To see this, observe that there must exist integers $\ell_1, \ell_2$ such that $j_1 \ell_2 - j_2 \ell_1 = 1$, as a consequence of our assumption that $\gcd(j_1, j_2) = 1$. Thus, we can define a new lattice basis as $\theta_1' = j_1 \theta_1 + j_2 \theta_2$ and $\theta'_2 = \ell_1 \theta_1 + \ell_2 \theta_2$. The Lagrangian \eqref{eq:Lmassmix}, written in the new basis, has the same form, with
\begin{align}
k'_1 &= \ell_2 k_1 - \ell_1 k_2, \nonumber \\
k'_2 &= -j_2 k_1 + j_1 k_2, \nonumber \\
\bm{K}' &= (\bm{M}^{-1})^{T} \bm{K} \bm{M}^{-1} \text{  where  } \bm{M} = \begin{pmatrix} j_1 & j_2 \\ \ell_1 & \ell_2 \end{pmatrix}.  \label{eq:latticebasistransform}
\end{align}
Here $\bm{K}$ denotes the kinetic matrix whose entries $K_{ij}$ appeared in \eqref{eq:Lmassmix}. Notice that the GL$(2,\bbZ)$ transformation maintains the quantization of couplings, $k'_i \in \bbZ$, as any lattice basis should. Without loss of generality, then, we can study the Lagrangian \eqref{eq:Lmassmix} in the special case that the potential depends only on $\theta'_1$. Let us do so, dropping the $'$ labels:
\be
{\cal L} = K_{ij} \partial_\mu \theta_i \partial^\mu \theta_j - V(\theta_1) + \frac{k_1 \theta_1 + k_2 \theta_2}{16\cpi^2} F_{\mu \nu} \tF^{\mu \nu}.
   \label{eq:Lmassmix1}
\ee
We could, equivalently, rewrite this in terms of two periodic, {\em dimensionful} axion fields $a_i$ with period $2\cpi F_i$, as in \eqref{eq:ashift}, with a dimensionless kinetic mixing parameter $\epsilon$:
\be
{\cal L} = \frac{1}{2} \sum_{i=1}^2 \partial_\mu a_i \partial^\mu a_i + \epsilon \partial_\mu a_1 \partial^\mu a_2 - V(a_1 / F_1) + \frac{1}{16\cpi^2} \left(k_1 \frac{a_1}{F_1} + k_2 \frac{a_2}{F_2}\right) F_{\mu \nu} \tF^{\mu \nu},
\ee
where
\be
F_i \df \sqrt{2 K_{ii}} \quad \text{and} \quad \epsilon \df \frac{K_{12}}{\sqrt{K_{11}K_{22}}}.
\ee

\subsubsection{Diagonalizing the propagating states}
\label{subsec:massdiag}

The Lagrangian \eqref{eq:Lmassmix1} clearly describes one massive propagating field, $\theta_1$, and another massless propagating field. For general $K_{ij}$, the massless field will be a general linear combination of $\theta_1$ and $\theta_2$, not necessarily aligned with any lattice vector. This means that it is not a periodic scalar, but rather winds around the torus in an irrational direction, never returning to its starting point. To identify this direction, we can diagonalize {\em both} the mass and the kinetic terms by performing a shift of the light field, i.e.~by defining
\begin{equation}
a_L \df a_2 + \epsilon a_1.
\end{equation} 
This resembles the familiar diagonalization of massive dark photons kinetically mixing with the massless ordinary photon \cite{Holdom:1985ag}, which was further discussed in \cite{Shiu:2013wxa}. To canonically normalize the independently propagating fields, we can further introduce a rescaled heavy field
\be
a_H \df \sqrt{1 - \epsilon^2} a_1.
\ee
In terms of $a_H$ and $a_L$, the Lagrangian takes the diagonalized form
\be
{\cal L} = \frac{1}{2} \partial_\mu a_L \partial^\mu a_L + \frac{1}{2} \partial_\mu a_H \partial^\mu a_H - V(a_H/f_H) + \frac{1}{16\cpi^2} \left(k_2 \frac{a_L}{f_L} + (k_1 - \rho k_2) \frac{a_H}{f_H} \right) F_{\mu\nu}\tF^{\mu \nu},   \label{eq:Lheavylightbasis}
\ee
where
\be
f_H \df \sqrt{1-\epsilon^2} F_1, \quad f_L \df F_2, \quad \text{and} \quad \rho \df \epsilon \frac{F_1}{F_2}.
\ee
We have denoted the suppression scale in the couplings by lowercase $f$ rather than capital $F$ to signal that, unlike in equation \eqref{eq:ashift}, they do not necessarily have an interpretation as the period of a compact boson. The quantity $\rho$ is essentially a measure of how misaligned the basis of propagating fields is with the lattice basis.

This form of the effective Lagrangian has been derived several times before, e.g., \cite{Babu:1994id, Higaki:2014qua, Agrawal:2017cmd}. However, there is an aspect of it that is, at first sight, puzzling and has not (to the best of our knowledge) been commented on. Specifically: the field $a_H$, being proportional to $\theta_1$, is a periodic scalar, yet its couplings to gauge fields depend on the (generically) irrational number $\rho$ and thus are not quantized. On the other hand, the field $a_L$ is {\em not} a periodic scalar, but its couplings to gauge fields are quantized (proportional to the integer $k_2$).

Should this bother us? Our argument that periodic scalars have quantized couplings was based on requiring that $\exp(\iu S)$ be gauge-invariant when the scalars are shifted. Because \eqref{eq:Lheavylightbasis} is fully equivalent to our manifestly gauge-invariant starting point \eqref{eq:Lmassmix}, it must be the case that $\exp(\iu S)$ is well-defined despite the non-quantized coupling of the periodic scalar $a_H$. The reason is that a gauge transformation $\theta_1 \mapsto \theta_1 + 2\cpi n$ does not {\em only} shift $a_H$, but also shifts $a_L$; our diagonalized Lagrangian \eqref{eq:Lheavylightbasis} is, as it must be, invariant under the gauge transformations
\begin{align}
a_H &\mapsto a_H + 2\cpi n_1 f_H,  \nonumber \\
a_L &\mapsto a_L + 2\cpi \left(n_2 + \rho n_1\right) f_L, \quad n_i \in \bbZ,
\end{align}
which simply reflect the coordinates of the lattice in our new, misaligned basis. The lack of quantization of the $a_H$ coupling leads to a shift in the action under a gauge transformation of $a_H$ that is precisely compensated by the corresponding, $\rho$-dependent shift of $a_L$ under the same gauge transformation. Everything is as it should be. However, one might wonder whether the lack of periodicity of $a_L$ means that we can integrate out $a_H$ and obtain a low-energy EFT of $a_L$ that lacks the constraints that usually come from periodicity. Given that our Lagrangian has quantized couplings of $a_L$ to $F \tF$, it does not seem to be so easy to escape the constraints of periodicity. In fact, the non-periodicity of $a_L$ is a red herring. Properly understood, the low-energy effective theory is a theory of a compact field, as we will now explain. 

\subsubsection{Periodicity in the low-energy EFT}
\label{subsec:periodicityEFT}

\begin{figure}[!h]
\centering
\includegraphics[width=0.4\textwidth]{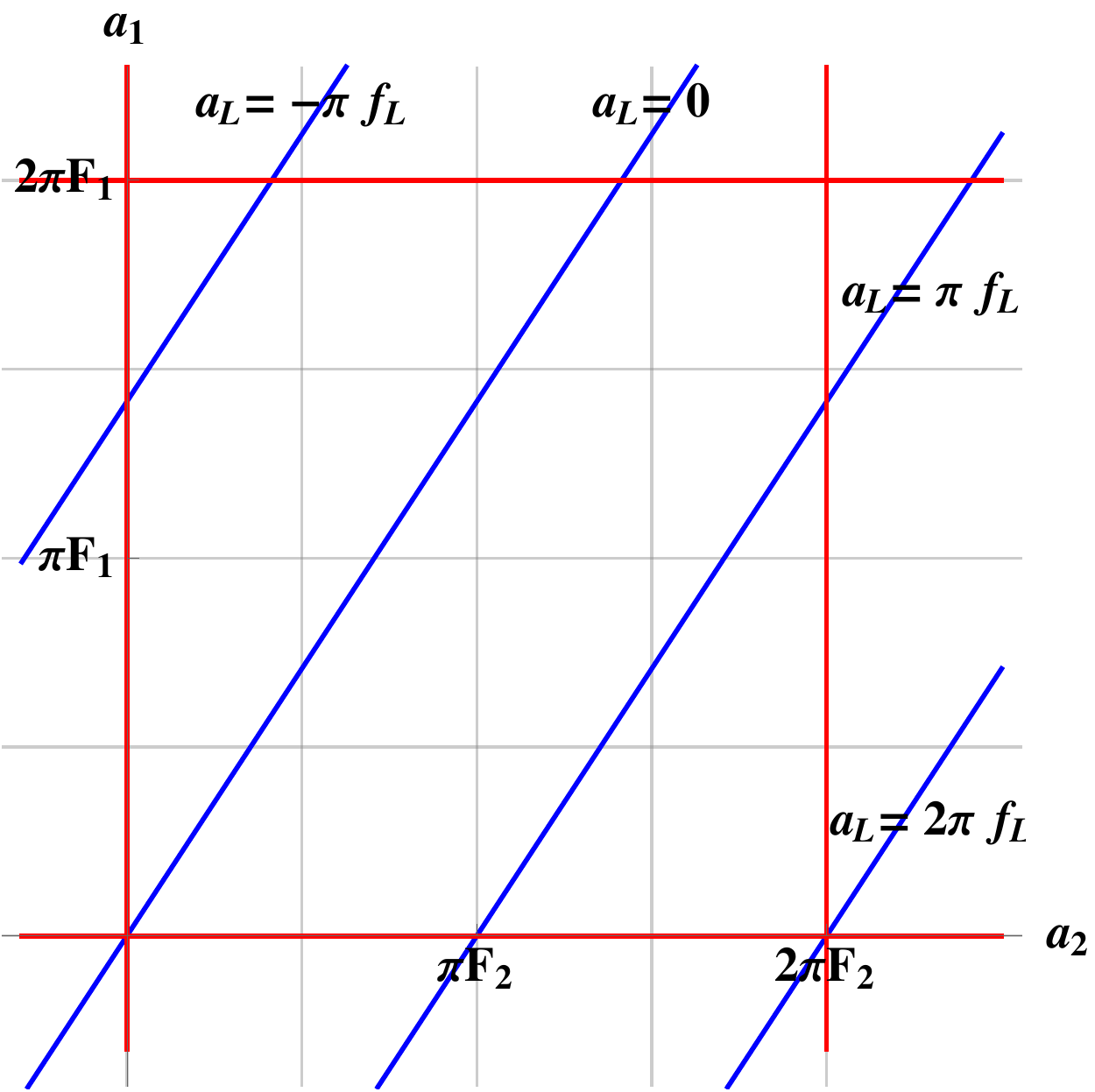}
\caption{The space of two axions $a_i$ with periods $2\cpi F_i$. The red lines indicate two periodically identified values of the two axions. The blue lines are contours of constant $a_L = a_2 + \epsilon a_1$, with $\epsilon = -\frac{1}{\sqrt{2}}$ chosen as an example. Notice that a $2\cpi$ shift of the heavy field $a_1$ shifts $a_L$ by an irrational amount, but a $2\cpi$ shift of $a_2$ at constant $a_1$ simply shifts $a_L \to a_L + 2\cpi f_L$. This is the gauge symmetry of the EFT along the flat valley minimizing the potential $V(a_1)$.}
\label{fig:twoaxionspace}
\end{figure}

We have noted that the light axion field $a_L$ is not a simple periodic field, but it still has quantized couplings. We can understand this better by examining the two-axion field space, as shown in Fig.~\ref{fig:twoaxionspace}. The field space consists of periodic variables $a_1$ and $a_2$, whereas when we diagonalize the kinetic terms we find a light field $a_L$ which is an irrational combination of the two, and which is constant on the blue diagonal lines in the plot. 

The potential $V(a_1)$ is independent of $a_2$, and hence constant along horizontal lines in this plot. This means that there is a flat valley along the horizontal red line at $a_1 = 0$, which is repeated at $a_1 = 2\cpi F_1$ and other gauge equivalent locations. The effective field theory of the light axion should be defined along this valley, since the field can move along it without incurring any potential energy cost. Notice that this statement is {\em independent} of the kinetic term for the axions, and in particular of the direction along which lines of constant $a_L$ are oriented. The gauge symmetry $a_1 \mapsto a_1 + 2\cpi F_1$ does not shift $a_L$ by a quantized multiple of $2\cpi f_L$. However, the only gauge symmetry that makes sense within the low-energy effective theory defined in a valley of fixed $a_1$, namely $a_2 \mapsto a_2 + 2\cpi F_2$, {\em does} shift $a_L$ by $2\cpi f_L$: it is the horizontal translation that takes, for instance, the diagonal at $a_L = 0$ to that at $a_L = 2\cpi f_L$. Furthermore, these facts are preserved by any lattice basis in which the potential depends only on $\theta_1$; we could send $\theta_2 \mapsto \theta'_2 \df \theta_2 + n \theta_1$, and in the $(\theta_1, \theta'_2)$ basis it is still true that $V$ is a function only of $\theta_1$ and that the couplings of $\theta'_2$ are quantized. The EFT with $\theta_1$ frozen at the minimum of its potential takes exactly the same form in the new basis.

From this point of view, there is very little mystery: the EFT is defined along the flat direction in field space, which is periodic. The couplings of the massless periodic axion should be quantized, and we have found that they are. A lesson to draw from this, which generalizes to other contexts, is that although diagonalizing the propagating states is a good way to proceed if you plan to do Feynman diagram calculations with multiple fields, it can be an unnecessarily confusing step in the process of understanding the correct way to think about the low-energy EFT.

A natural, straightforward approach to understanding the low-energy EFT is to obtain a theory of the periodic field $\theta_2$ by directly integrating out $\theta_1$ using its equation of motion,
\begin{equation}
2 K_{11} \Box \theta_1 + 2 K_{12} \Box \theta_2 + \mu^4 \theta_1 - \frac{k_1}{16\cpi^2} F_{\mu \nu} \tF^{\mu \nu} + \ldots = 0,
\end{equation} 
where the omitted terms arise from higher orders in the Taylor expansion of $V(\theta_1)$. Solving this equation reveals that
\begin{align}
\theta_1 &= - \frac{2 K_{12} \Box \theta_2}{\mu^4} + \frac{4 K_{11} K_{12} \Box^2 \theta_2}{\mu^8} + \frac{k_1}{16\cpi^2 \mu^4} F_{\mu \nu}\tF^{\mu \nu} + \ldots \nonumber \\
&=\frac{1}{F_1} \left[ -\frac{\epsilon}{m_1^2} \Box a_2 + \frac{\epsilon}{m_1^4} \Box^2 a_2 + \frac{k_1}{16\cpi^2 m_1^2 F_1} F_{\mu \nu} \tF^{\mu \nu} + \ldots \right],   \label{eq:theta1EFT}
\end{align}
where in the second line we have rescaled to dimensionful fields and made the replacement $\mu^4 = m_1^2 F_1^2$. This makes it apparent that we could equally well obtain such an expansion by working with Feynman diagrams defined in terms of the fields $\theta_{1,2}$ rather than the diagonalized fields. The kinetic mixing is then an insertion proportional to $\epsilon \Box$ that flips a $\theta_2$ propagator to a $\theta_1$ propagator or vice versa, and leads to the $\Box a_2$ terms in the above equation.

In summary: the EFT of the light field is a theory of an axion $\theta_2$, with couplings to gauge fields quantized as expected given its periodicity. All of the effects of kinetic mixing with the heavy field are encoded in manifestly shift symmetry-preserving derivative interactions.

\subsection{An $N$-axion generalization}
\label{subsec:Naxiongen}

Above, we saw that if we had two axions $(\theta_1, \theta_2)$ and a potential depending on one linear combination of the two, we could change to a new lattice basis in which the potential is independent of the periodic axion $\theta'_2$. This allows us to integrate out the heavy field and obtain a theory of only the compact axion $\theta'_2$. It is natural to generalize this to the case of $N$ axions $(\theta_1, \ldots, \theta_N)$ with period $2\cpi$ as follows: suppose that we have a potential that depends on $k$ independent linear combinations of the $N$ axions and respects their periodicity, 
\be
V = V(\vartheta_1, \ldots, \vartheta_k) \quad \text{where} \quad \vartheta_i = \sum_{j=1}^N Q_{ij} \theta_j, \quad Q_{ij} \in \bbZ.
\ee
Then we claim that there is a new lattice basis, $\theta'_1, \ldots, \theta'_N$, in which the potential has the form $V(\theta'_1, \ldots, \theta'_k)$ and is independent of $\theta'_{k+1}, \ldots, \theta'_N$. Hence, we can integrate out the massive modes $\theta'_1, \ldots, \theta'_k$ to obtain an effective field theory of the $N - k$ massless, $2\cpi$-periodic axions $\theta'_{k+1}, \ldots, \theta'_N$.

This fact follows from the existence of the {\em Smith normal form} \cite{smith1862systems} for matrices over a principal ideal domain (such as the integers): given the $k \times n$ integer matrix $\bm{Q}$ with entries $Q_{ij}$, there exist integer matrices $\bm{S} \in \text{GL}(k, \bbZ), \bm{T} \in \text{GL}(n, \bbZ)$ such that
\be
\bm{R} \df \bm{S} \bm{Q} \bm{T} = \begin{pmatrix} r_1 & 0 & \cdots & 0 & 0 & \cdots & 0 \\
                                                                                   0 & r_2 & \cdots & 0 & 0 & \cdots & 0 \\
                                                                                   0 & 0 & \ddots & 0 & 0 & \cdots & 0 \\
                                                                                   0 & 0 & \cdots & r_k & 0 & \cdots & 0 \end{pmatrix}  , \quad r_i \in \bbZ. 
\ee
(The Smith decomposition also implies that we can arrange that $r_i$ divides $r_{i+1}$, but we will not need this.) The definition of GL$(m,\bbZ)$ means that $\bm{S}$ and $\bm{T}$ are invertible and their inverses have integer entries.

In our original basis, the span of the rows of $\bm{Q}$ defines the subspace of axions that obtain a potential. We can change to a new lattice basis by defining
\be
\theta'_i = \sum_{j=1}^N T^{-1}_{ij} \theta_j,
\ee
where $T^{-1}_{ij}$ are the entries in $\bm{T}^{-1}$. In terms of this basis, the potential depends on the span of the rows of the matrix $\bm{Q}\bm{T} = \bm{S}^{-1} \bm{R}$. We can read off immediately that the span of the rows of $\bm{R}$ contains only linear combinations of the first $k$ basis vectors in the $\theta'_i$ basis. The form of $\bm{R}$ together with invertibility of $\bm{S}^{-1}$ guarantees that the rows of $\bm{S}^{-1} \bm{R}$ span the same subspace. Hence the potential is independent of $(\theta'_{k+1}, \ldots, \theta'_N)$.

This shows that our discussion of the 2-axion case can be fully generalized to $N$ axions. When a potential gives a mass to $k < N$ axions, we can always find a lattice basis where it is manifest that $N - k$ axions with period $2\cpi$ are flat directions. By the usual logic of effective field theory, then, we can integrate out all of the heavy axions, and obtain a theory of the $N-k$ light axions that respects all of the expected quantization rules for axion couplings. Any kinetic mixing with heavy axions, upon integrating them out, will produce only shift-symmetric terms involving $\Box$ acting on light axions, as we saw above.

\subsection{Light axion obtaining a mass}
 \label{subsec:lightaxionmass}

So far we have discussed only cases in which light axions remain exactly massless, and have found that they are periodic fields with exactly quantized couplings. The quantization of axion couplings can be violated once the axions obtain a mass. One straightforward way to see this is by noting that within the effective field theory, we can use equations of motion to make the replacement
\be
\Box a_L \mapsto -\frac{\partial V(a_L)}{\partial a_L} \approx -m_L^2 a_L + \ldots, \label{eq:aLeom}
\ee
which exchanges a term that is manifestly invariant under continuous shift symmetries of $\theta_2$ with one that is not. Although the linear term coupling $a_L$ to $F \tF$ is not necessarily quantized, if we keep higher-order terms in $a_L$ this replacement does preserve the discrete shift symmetry \eqref{eq:ashift} because $V(a_L)$ is a periodic function. One could also see this effect from the Feynman diagram approach; an external, on-shell light axion of mass $m_L$ that kinetically mixes with the off-shell propagator of a heavier axion of mass $m_H$ will obtain an insertion proportional to $\epsilon p^2 = \epsilon m_L^2$ followed by a propagator factor of $1/(m_H^2 - m_L^2)$, which agrees with the EFT result obtained by integrating out $\theta_1$ using \eqref{eq:theta1EFT} and expanding order-by-order in $m_L^2/m_H^2 \ll 1$. Thus, the couplings of a massive axion field are not quantized, but to the extent that the mass of the axion is much smaller than all other mass scales in the problem, we expect the deviations from coupling quantization to be small.

It is instructive to compare this to the familiar non-quantized shift of the axion coupling to photons via its mixing with the neutral pion. As explained in \cite{Agrawal:2017cmd}, this does not violate the shift symmetry \eqref{eq:ashift} of the axion because it is part of a set of terms that resum to a periodic function, similar to \eqref{eq:aLeom}. Furthermore, the effect is suppressed by $m_a^2$, and is large only because the axion mass arises from the same strong dynamics as the pion mass, so that $m_a^2 F_a^2 \sim m_\pi^2 F_\pi^2$. In other words, in this case, the suppression factor $m_a^2/m_\pi^2$ that we have argued to exist on general EFT grounds is compensated by an enhancement factor of $F_a^2/F_\pi^2$. (One could, in principle, perform a field redefinition to discuss this example in the language of kinetic mixing rather than mass mixing, although because the kinetic mixing would then be nearly maximal, this is not a very useful viewpoint to take.) This example shows that some caution is in order when asserting that effective field theories of very light axions are expected to contain quantized couplings to gauge fields. On the other hand, it also reveals that one needs rather special circumstances to obtain a very large violation of this expectation, as arises when multiple periodic scalars obtain mass simultaneously from the same dynamics, as in QCD confinement.

\section{Mixing with a Heavier Axion Eaten by a Spin-1 Field}
\label{sec:spin1mix}

As our next example, we again consider a theory with two axions, but with a linear combination obtaining a mass in a different way: by being eaten by a massive, spin-1 gauge field through the Higgs or St{\"u}ckelberg mechanism \cite{Cheng:2001ys, Anastasopoulos:2006cz}. This type of theory has been considered in great detail in \cite{Shiu:2015uva, Shiu:2015xda}. A version of it in a Randall-Sundrum scenario was recently discussed in \cite{Fonseca:2019aux}. In this scenario, we will again see intermediate results that seem to break the expected connection between periodic scalar fields and quantized couplings. In this case, the pattern will be reversed from what we observed in \S\ref{subsec:massdiag}: the heavy axion will be a {\em non}-periodic field, but will have quantized couplings; on the other hand, the light axion will be a periodic field, but will have non-quantized couplings. These non-quantized couplings have led to earlier claims that super-Planckian field ranges can be obtained in models of this type \cite{Shiu:2015uva, Shiu:2015xda, Fonseca:2019aux}. However, our results do not support such claims. Once again, a careful assessment of the underlying gauge invariance of the theory will show that the proper understanding of the low-energy EFT is that of a periodic field with quantized couplings, despite initial appearances. Our conclusions are in accord with those of \cite{Choi:2019ahy}, which appeared while this paper was being completed.

\subsection{Diagonalizing the propagating states}
\label{subsec:spin1diag}

Let's begin by looking at an effective theory with two axions, one combination of which is eaten to provide a mass to a spin-1 field via the St{\"u}ckelberg mechanism. (It is possible to obtain this effective theory as a limiting case of a Higgs mechanism, so we expect our remarks to apply to both scenarios.) For the time being, we will neglect kinetic mixing, as the points we wish to illustrate do not depend on it.
We begin by considering the action
 \be
\sum_{i=1}^2 \frac{1}{2} F_i^2 (\partial_\mu \theta_i - q_i A_\mu)^2  - \frac{1}{4e^2} F_{\mu \nu}F^{\mu \nu} - \frac{1}{4g^2} G_{\mu \nu}G^{\mu \nu} + \frac{k_1 \theta_1 + k_2 \theta_2}{16\cpi^2} G_{\mu \nu}\tG^{\mu \nu} + {\cal L}_\text{con},   \label{eq:LStu}
\ee
where $F_{\mu \nu} = \partial_\mu A_\nu - \partial_\nu A_\mu$ is the field strength of the massive gauge field, whereas $G_{\mu \nu}$ is the field strength of a different, massless gauge field $G_\mu$. We are interested in the quantization of the $\theta G \tG$ coupling for the light axion. The term ${\cal L}_\text{con}$ denotes additional couplings that, in some cases, may be necessary for consistency of the theory. We will discuss these couplings in more detail below.

The subtleties in this case, compared to our previous case, arise because we now must ensure invariance under {\em three} different gauge transformations that shift the axions. These are the two discrete shift symmetries $\theta_i \mapsto \theta_i + 2\cpi$ associated with the periodicities of the axions, together with a continuous shift symmetry associated with the U(1) group gauged by $A_\mu$:
\begin{align}
A_\mu \mapsto A_\mu + \partial_\mu \alpha,\quad \theta_1 \mapsto \theta_1 + q_1 \alpha,\quad \theta_2 \mapsto \theta_2 + q_2 \alpha.  \label{eq:U1shifts}
\end{align}
When studying the theory on a spacetime of nontrivial topology, $\E^{\iu \alpha(x)} \in {\rm U}(1)$ must be well-defined (single-valued) but $\alpha$ itself need not be. Because $\E^{\iu \theta_i(x)}$ must also be single-valued, we see that the gauge transformation \eqref{eq:U1shifts} makes sense only if $q_1, q_2 \in \bbZ$. This is consistent with our expectations if the axions $\theta_i$ arise as phases of complex fields of charge $q_i$ that obtain a vacuum expectation value, in the case that \eqref{eq:LStu} arises as a limit of the Higgs mechanism.

Consistency of the theory under the axion shift symmetries imposes that $k_1, k_2 \in \bbZ$ in the absence of additional interactions, just as in our earlier discussions. However, notice that in general the $\theta G \tG$ terms are {\em not} invariant under the U(1) gauge transformation \eqref{eq:U1shifts}, which shifts the Lagrangian by
\be
\delta_\alpha {\cal L}_{\theta G \tG} = \frac{k_1 q_1 + k_2 q_2}{16\cpi^2} \alpha G_{\mu \nu} \tG^{\mu \nu}. \label{eq:gaugemismatch}
\ee
Because $\alpha$ is a continuous quantity, the theory only respects the U(1) gauge symmetry if $k_1 q_1 + k_2 q_2 = 0$. Otherwise, it is necessary to add additional terms, indicated by ${\cal L}_\text{con}$ above, which are not gauge invariant on their own but which serve to cancel the gauge variation \eqref{eq:gaugemismatch}. Such terms could arise from fermions that carry $G$ charge and transform anomalously under the U(1), or from generalized Chern-Simons terms proportional to $A_\mu K^\mu$ where $\partial_\mu K^\mu = G_{\mu \nu} \tG^{\mu \nu}$ \cite{Anastasopoulos:2006cz, Shiu:2015uva, Shiu:2015xda}.

For the moment, let us leave ${\cal L}_\text{con}$ unspecified and proceed to diagonalize the propagating states in \eqref{eq:LStu}. We can change basis to diagonalize the kinetic terms,
\be
{\cal L} \supset \frac{1}{2} m_A^2 (\partial_\mu a_H - A_\mu)(\partial^\mu a_H - A^\mu) + \frac{1}{2} m_A^2 \partial_\mu a_L \partial^\mu a_L,
\ee
where the ``heavy axion,'' the linear combination eaten by the U(1) gauge boson with mass $m_A$, is
\begin{align}
a_H \df \frac{1}{m_A^2} \left[F_1^2 q_1 \theta_1 + F_2^2 q_2 \theta_2\right], \quad \text{where} \quad m_A^2 \df F_1^2 q_1^2 + F_2^2 q_2^2.
\end{align}
The orthogonal light combination is
\be
a_L \df \frac{F_1 F_2}{m_A^2} \left(q_2 \theta_1 - q_1 \theta_2\right).  \label{eq:aL}
\ee
The proportionality of $a_L$ to an integer linear combination of our original axion fields is no accident; it guarantees that $a_L$ is invariant under the U(1) gauge transformation \eqref{eq:U1shifts}. In this basis, the couplings of the propagating axion eigenstates to the gauge field $G$ are
\be
{\cal L} \supset \frac{1}{16\cpi^2} \left[\left(k_1 q_2 \frac{F_2}{F_1} - k_2 q_1 \frac{F_1}{F_2}\right) a_L + \left(k_1 q_1 + k_2 q_2\right) a_H\right] G_{\mu \nu} \tG^{\mu \nu}.
\ee
These results have already been obtained in \cite{Shiu:2015uva, Shiu:2015xda}, but let us discuss them from the point of view of gauge invariance, periodicity, and quantized couplings. The fact that the coupling of $a_H$ is proportional to $k_1 q_1 + k_2 q_2$ follows from \eqref{eq:gaugemismatch}: when the Lagrangian is gauge invariant without further contributions, i.e., when ${\cal L}_\text{con} = 0$, U(1) gauge anomaly cancellation demands that the linear combination of axions that transforms under U(1) decouples from the other gauge fields.

Recall that in \S\ref{subsec:massdiag}, we faced a puzzle: after diagonalizing the mass and kinetic mixings, we found a heavy propagating axion mass eigenstate that was a periodic scalar and yet had non-quantized couplings, and a massless light axion that was not a periodic scalar and yet had quantized couplings. Here we seem to see exactly the opposite situation: the heavy linear combination $a_H$ is not a periodic scalar, and yet its coupling to $G \tG$ is proportional to the integer $k_1 q_1 + k_2 q_2$.\footnote{This is a bit of an overstatement: in the presence of ${\cal L}_\text{con}$, as noted above, the $k_i$ need not be integers, whereas in the absence of ${\cal L}_\text{con}$, this coupling is not just any integer, but zero.} On the other hand, the light axion $a_L$ is periodic; from \eqref{eq:aL} we can see that under a general shift of the underlying fields $\theta_i \mapsto \theta_i + 2\cpi n_i$, the shift of $a_L$ is proportional to the integer $q_2 n_1 - q_1 n_2$. Thus, $a_L$ is periodic, with minimal period given by the identification
\be
a_L \cong a_L + 2\cpi \frac{F_1 F_2}{m_A^2}  \gcd(q_1,q_2).
\ee
The puzzle is that despite this periodicity, the couplings of $a_L$ do not seem to be quantized, as the $G \tG$ coupling depends not only on the integer charges $k_i, q_i$, but also on the ratio of decay constants $F_1 / F_2$. In particular, if we define a scalar field $\theta_L$ of period $2\cpi$ by rescaling $a_L$,
\be
\theta_L \df \frac{1}{\gcd(q_1,q_2)} (q_2 \theta_1 - q_1 \theta_2),
\ee
its coupling to $G \tG$ is given by
\be
\frac{1}{16\cpi^2} \left[\gcd(q_1,q_2) \left(\frac{k_1 q_2 F_2^2 - k_2 q_1 F_1^2}{q_1^2 F_1^2 + q_2^2 F_2^2}\right)\right] \theta_L G_{\mu \nu} \tG^{\mu \nu},   \label{eq:aLGtGcoupling}
\ee
where the term in brackets is, in general, {\em not} an integer. This appears to contradict the basic periodicity \eqref{eq:LthetaFF} of a $2\cpi$-periodic axion.

We emphasize that the periodicity of $\theta_L$ is entirely determined by the original periodic lattice of identifications of $(\theta_1, \theta_2)$ together with the charges $q_i$, which specify which linear combination of the fields remains uneaten. We can write the kinetic term of $\theta_L$ as 
\be
\frac{1}{2} F_L^2 \partial_\mu \theta_L \partial^\mu \theta_L, \quad \text{where} \quad F_L = 
\frac{F_1 F_2 \gcd(q_1, q_2)}{m_A}.   \label{eq:thetaLnorm}
\ee
The scale $F_L$ is the most natural definition of the ``decay constant'' of the light axion, and determines the units in which the couplings of the canonically normalized axion to $G \tG$ are expected to be quantized as well as the expected field range when an axion potential is generated. This should be contrasted with the approach of \cite{Shiu:2015uva, Shiu:2015xda, Fonseca:2019aux}, which defines an effective decay constant $F_\text{eff}$ which is inversely proportional to the factor in brackets in \eqref{eq:aLGtGcoupling}. In those studies, a small value of the factor $k_1 q_2 F_2^2 - k_2 q_1 F_1^2$ is argued to suppress the coupling and lead to a trans-Planckian $F_\text{eff}$. While it is interesting that the coupling in \eqref{eq:aLGtGcoupling} allows for a very large $F_\text{eff}$ defined in this way, the fact that it is not related to the period $2\cpi F_L$ appearing in \eqref{eq:thetaLnorm} should give us pause. In fact, the axion decay constant as we have defined it can only be smaller than the decay constants we started with:
\be
F_L < \min(F_1, F_2).
\ee 
How do we reconcile this with claims of large $F_\text{eff}$ extracted from \eqref{eq:aLGtGcoupling}?

We have already laid the groundwork for the resolution of this puzzle: because the Lagrangian \eqref{eq:LStu} is {\em not}, in general, gauge-invariant in the absence of additional terms ${\cal L}_\text{con}$, we should not be surprised that it violates the expected periodicity properties. The physical coupling of an axion to gauge fields is quantized in units that allow us to read off the maximum field range of the axion potential, but in theories with ${\cal L}_\text{con} \neq 0$, Lagrangian couplings like that in \eqref{eq:aLGtGcoupling} do not determine the full amplitude, and consequently we do not expect that the scale $F_{\rm eff}$ extracted from such a term is related to a physical field range. The discussion in the introduction makes this clear: if we read off $F_{\rm eff} \propto 1/c_F$ from \eqref{eq:thetaFFfermions}, then because $c_F$ shifts as in \eqref{eq:fieldredefcoupling} under a field redefinition, we could obtain absolutely any value of $F_{\rm eff}$ by parametrizing our fields in a different way. The physical amplitude which is quantized, in the presence of anomalous fermions, depends on a combination of terms like \eqref{eq:invariantquantization}. Only by first redefining the fermions to set $c_m = 0$ (which, in this context, is the meaning of ${\cal L}_\text{con} = 0$) do we obtain quantized $c_F$, at which point we can read off the axion periodicity from this coupling. Hence, we cannot, in general, analyze the periodicity constraints on the effective action of $\theta_L$ without specifying the terms ${\cal L}_\text{con}$, which we expect will always resolve the puzzle. The only case in which we can directly resolve the puzzle is in the case when it is consistent to set ${\cal L}_\text{con} = 0$ because $\delta_\alpha {\cal L}_{\theta G \tG}$ in \eqref{eq:gaugemismatch} is identically zero, i.e., the case $k_1 q_1 + k_2 q_2 = 0$. In this case, the bracketed factor in \eqref{eq:aLGtGcoupling} reduces to
\be
\left[\gcd(q_1,q_2) \left(\frac{k_1 q_2 F_2^2 - k_2 q_1 F_1^2}{q_1^2 F_1^2 + q_2^2 F_2^2}\right)\right]  \mapsto -\frac{k_2}{q_1} \gcd(q_1,q_2) \in \bbZ.  \label{eq:divisibility}
\ee
To justify the claim that this is an integer: given that $k_1 q_1 = -k_2 q_2$, it follows that $q_1 | (k_2 q_2)$. In order for this to be true, $q_1/\gcd(q_1, q_2)$ must divide $k_2$.

So far we have assumed the light axion to be exactly massless, and found that it has exactly quantized couplings. We could also consider a theory which has a potential that provides a mass for the light axion well below the mass of the heavy spin-1 field. Just as we discussed in \S\ref{subsec:lightaxionmass}, the effective field theory of the light axion allows for terms proportional to $\Box \theta_L$ which, upon making use of the equations of motion, can appear as effectively non-quantized couplings proportional to the light axion mass squared.

Summing up: when we give one linear combination of the axions a mass through the Higgs or St{\"u}ckelberg mechanism, the massless light axion is a periodic field, with smaller field range than our initial axions. In the case that the Lagrangian we have studied is gauge invariant in its own right, we have shown that the couplings of this periodic field are quantized, just as we expect them to be. This is as it must be; if we integrate out the heavy fields, we obtain an effective field theory of a periodic axion, with all of the constraints that this entails. Nonetheless, to illustrate the point more generally, let us look at at an example in which ${\cal L}_\text{con} \neq 0$. Specifically, we will consider a theory in which light fermions cancel the gauge variation \eqref{eq:gaugemismatch}.

\subsection{Analyzing a 4d UV completion}

To clarify the physics, it is useful to consider an explicit, 4d UV completion of the effective Lagrangian \eqref{eq:LStu} in which the massive gauge field obtains a mass from the Higgs mechanism, and fermion fields supply a non-vanishing contribution to ${\cal L}_\text{con}$. The goal of this model is simply to show a consistent example that generates the effective theory we are interested in, in which we can explicitly calculate the interactions and understand how the constraints of axion periodicity are respected. This model is not meant to be natural or aesthetically appealing, just to illustrate some points about the physics of axions. For this reason, we will freely assume hierarchies in the dimensionless couplings, and invoke global symmetries that are not necessarily accidental, with no need for further explanation.

In this model, the axions $\theta_{1,2}$ arise from the phases of two complex scalars $\phi_{1,2}$ with U(1) gauge charges $q_{1,2}$. We also consider an SU($N$) gauge group that will provide the $G\tG$ couplings we are interested in. Each of the scalars will provide Dirac masses to some fermions $Q, \tQ$ which transform in non-trivial, conjugate SU($N$) representations, so that from the SU($N$) point of view the theory is not chiral. However, these fields will have chiral couplings to U(1): $Q$ carries charge and $\tQ$ does not, or vice versa. To cancel the U(1)$^3$ and mixed U(1)--gravitational anomalies, we also introduce fermions $L, \tL$ that have the opposite U(1) charge assignments but do {\em not} interact with SU($N$) gauge fields (though they come in the appropriate number of copies to compensate for the anomalies of the $Q, \tQ$ fields). By construction, this theory has no SU($N$)$^3$, U(1)$^3$, or mixed gravitational anomaly, but the $\mathrm{SU}(N)^2\mathrm{U}(1)$ mixed anomaly still imposes a nontrivial constraint on the representations and charged assignments, to which we will return shortly. The field content of this model is summarized in Table~\ref{table:4dfields}.

\begin{table}[!h]
\begin{center}
{\tabulinesep=1.2mm
\begin{tabu}{l|l|l|l|l|l|l|}
                & $\phi_1$ & $\phi_2$ & $Q_{1i}$ ~\hfill~$\tQ_{1i}$      & $L_{1ik}$~\hfill~$\tL_{1ik}$ & $Q_{2j}$~ \hfill~$\tQ_{2j}$      & $L_{2jk}$~\hfill~$\tL_{2jk}$ \\ \hline
U(1)$_\text{gauge}$  & $q_1$    & $q_2$    & $q_1$\hfill $0$                        & $-q_1$\hfill $0$                     & $0$ \hfill $q_2$  & $0$ \hfill $-q_2$                     \\ \hline
SU($N$)         & 1        & 1        & $R_{1i}$ \hfill $\overline{R}_{1i}$ & $1$\hfill$1$                        & $R_{2j}$ \hfill $\overline{R}_{2j}$ & $1$\hfill$1$                         \\ \hline
U(1)$_\text{global}$ & $1$        & $0$        & $1$\hfill$0$                             & $0$\hfill$-1$                       & $0$\hfill$0$                             & $0$\hfill$0$                   \\ \hline
$N_\text{copies}$ & 1 & 1 & 1 & $\dim(R_{1i})$ & 1 & $\dim(R_{2j})$ \\ \hline
\end{tabu}}
\caption{Matter field content in a potential UV completion of the two axion model. The integers $i \in \{1, \ldots, N_1\}$ and $j \in \{1, \ldots, N_2\}$ label the set of fields, while the subscripts $1$ and $2$ signal which Yukawa couplings provide mass to the fields, e.g., $\phi_1 Q_{1i} \tQ_{1i}$. The full set of Yukawa couplings is displayed in \eqref{eqn:Yukawa}. The $\phi$ fields are scalars, whereas the $Q, \tQ, L, \tL$ fields are all left-handed Weyl fermions. The $L, \tL$ fields come in multiple copies, $k \in \{1, \ldots, N_\text{copies}\}$, to ensure anomaly cancellation. With these charge assignments, the only anomaly cancellation condition that must be explicitly checked is the $\mathrm{SU}(N)^2\mathrm{U}(1)$ anomaly.}
\label{table:4dfields} 
\end{center}
\end{table}

\begin{figure}[!h]
\centering
\includegraphics[width=1.0\textwidth]{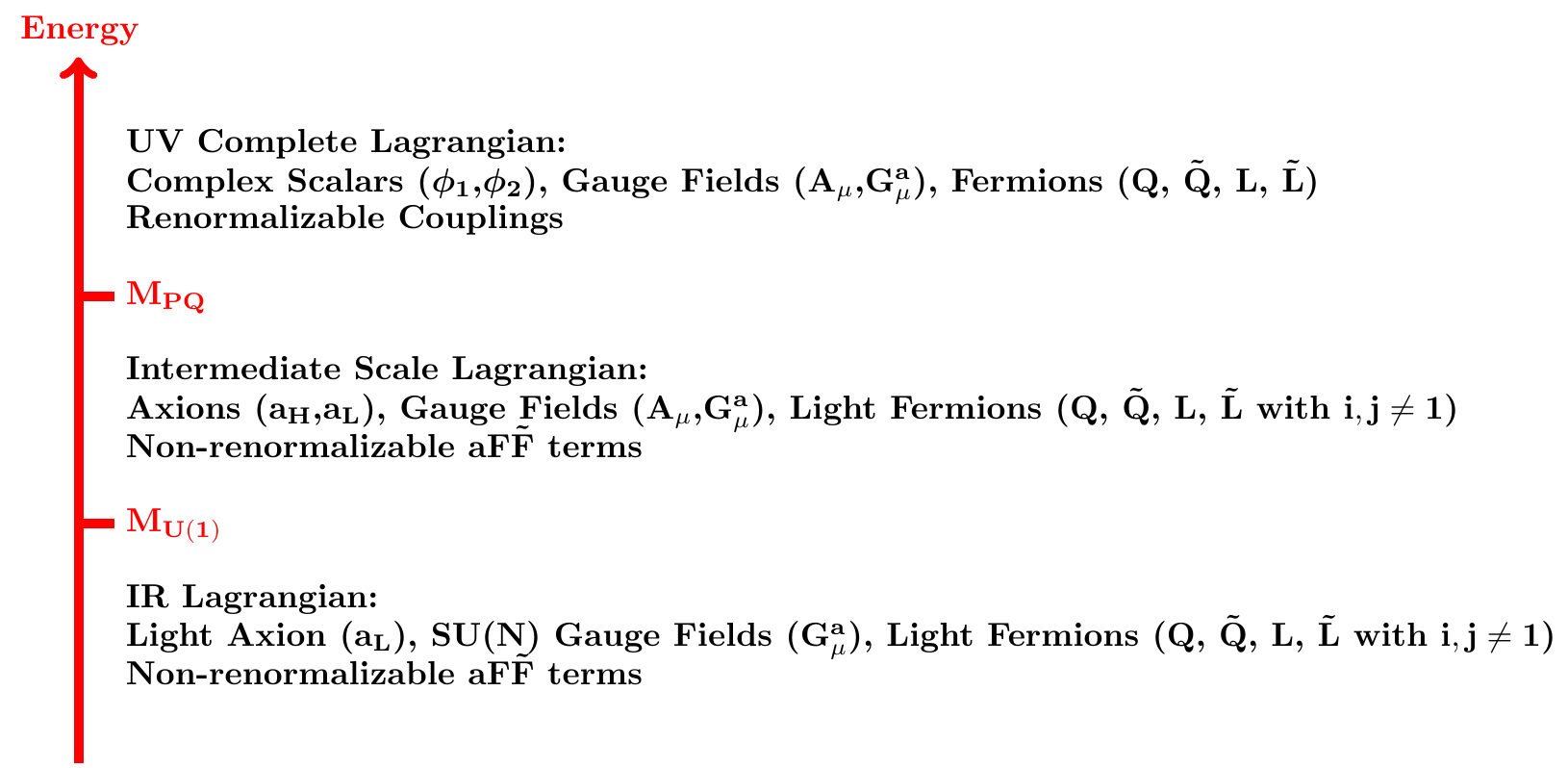}
\caption{Schematic of important energy scales and effective field theories obtained from our UV completion. After integrating out the radial modes of the scalars and the heavy fermions at the Peccei-Quinn scale $M_{\rm PQ}$, we obtain an effective field theory of two axions, one eaten by a spin-1 field, as considered in \S\ref{subsec:spin1diag}. Below the mass scale $M_{\text{U(1)}}$ of the spin-1 field, an effective field theory of a single light axion is obtained. We show that this axion is a periodic field with quantized couplings.}
\label{fig:energyscales}
\end{figure}

The Lagrangian \eqref{eq:LStu} can be obtained in a decoupling limit of this model. We begin with the complete theory, including Yukawa couplings
\begin{equation}
\Lag_\text{Yuk} = \sum_i \left( y_{1Qi} \phi_1^\dag  Q_{1i} \tQ_{1i} +  y_{1Li} \phi_1 L_{1ik} \tL_{1ik}\right) + \sum_j \left(y_{2Qi} \phi_2^\dag Q_{2j} \tQ_{2j} + y_{2Lj} \phi_2 L_{2ik} \tL_{2ik} \right) + \text{h.c.},
\label{eqn:Yukawa}
\end{equation}
where sums over the copies $k$ of the $L$ fields are implicit. To generate an effective Lagrangian of the form \eqref{eq:LStu}, we suppose that there is a hierarchy among the Yukawa couplings so that some are much larger than others. Then below the symmetry breaking scale, we can integrate out the heavy fermions. In general, integrating out a term of the form $m(\phi) \Psi \widetilde{\Psi} + \text{h.c.}$ produces a term of the form
\begin{equation}
\Delta \Lag = \frac{2\mu(R_\Psi)}{32 \cpi^2} \arg(m) G_{\mu \nu}^a \tG^{a \mu \nu},
\end{equation}
where $\mu(R_\Psi)$ is the Dynkin index of the representation of $\Psi$ under the group $G$. For concreteness, let us suppose that the fields with $i = 1$ and $j = 1$ are relatively heavy, whereas all of the others are much lighter (i.e., have much smaller Yukawa couplings to $\phi_1, \phi_2$). We further assume that the U(1) gauge coupling $e$ is small enough that we can integrate out the heavy fermions {\em without} integrating out the massive gauge field, i.e., $eq_{1,2} \ll y_{1Q1}, y_{1L1}, y_{2Q1}, y_{2L1}$. We further assume that the fields $\phi_{1,2}$ have a symmetry breaking potential which does not mix them, e.g.,
\be
V_\text{SSB} = \frac{\lambda_1}{4} \left(|\phi_1|^2 - v_1^2\right)^2 + \frac{\lambda_2}{4} \left(|\phi_2|^2 - v_2^2\right)^2.
\ee
The structure of this potential ensures that, when we turn off the U(1) gauge interaction, we have two distinct Nambu-Goldstone bosons $\theta_{1,2}$ which are the phases of $\phi_{1,2}$ respectively. An example of a U(1) global symmetry charge assignment that can be responsible for protecting the uneaten Nambu-Goldstone boson is given in the ``U(1)$_\text{global}$'' row of Table~\ref{table:4dfields}. We further assume that the radial modes of the $\phi$ fields are sufficiently heavy that we can integrate them out, i.e., $e q_{1,2} \ll \sqrt{\lambda_{1,2}}$. The choice of which fields to integrate out is not unique, but making this arbitrary choice suffices to illustrate our main points. We illustrate the various interesting ranges of energies, and corresponding effective field theories, in this model in Fig.~\ref{fig:energyscales}.

Integrating out the heavy SU($N$)-charged fermions $Q_{11}, \tQ_{11}, Q_{21}, \tQ_{21}$ will generate couplings
\be
\Lag_{\theta G \tG} = -\frac{1}{16\cpi^2} \left[\mu(R_{11}) \theta_1 + \mu(R_{21}) \theta_2\right] G^a_{\mu \nu} \tG^{a\mu \nu}.
\ee
We follow the standard convention in particle physics that the Dynkin index of the fundamental representation of SU($N$) is $\mu(\Box) = 1/2$, in which case the Dynkin index of any representation $R$ satisfies $2\mu(R) \in \bbZ$, which shows that the couplings of $\theta_{1,2}$ are quantized in the way that we expect. Changing basis as described in \S\ref{subsec:spin1diag}, this includes a coupling of the light axion $\theta_L$ of the form
\be
-\frac{\gcd(q_1,q_2)}{16\cpi^2} \frac{1}{m_A^2} \left[\mu(R_{11}) q_2 F_2^2 - \mu(R_{21}) q_1 F_1^2\right] \theta_L G^a_{\mu \nu} \tG^{a\mu \nu}.  \label{eq:thetaLintermed}
\ee
While it appears that we can choose this to be as small as we like by carefully choosing representations to impose relations among the $q_l$ and $\mu(R_{mn})$, we have not yet taken into account gauge invariance. The condition for $\mathrm{SU}(N)^2\mathrm{U}(1)$ anomaly cancellation, given the field content in Table~\ref{table:4dfields}, is
\be
\sum_{i = 1}^{N_1} \mu(R_{1i}) q_1 + \sum_{j=1}^{N_2} \mu(R_{2j}) q_2 = 0.   \label{eq:SUN2U1cancel}
\ee
We can use this condition to eliminate $\mu(R_{11})$ from \eqref{eq:thetaLintermed}, obtaining a coupling
\begin{align}
{\cal L}_{\theta_L G \tG} =& -\frac{\gcd(q_1,q_2)}{16\cpi^2} \frac{1}{m_A^2} \left[-\sum_{i = 2}^{N_1} \mu(R_{1i})q_2 F_2^2 - \sum_{j=1}^{N_2} \mu(R_{2j}) \frac{q_2^2}{q_1} F_2^2 - \mu(R_{21}) q_1 F_1^2\right]  \theta_L G^a_{\mu \nu} \tG^{a\mu \nu} \nonumber \\
= & \frac{\gcd(q_1,q_2)}{16\cpi^2} \left[\frac{\mu(R_{21})}{q_1} +\sum_{i = 2}^{N_1} \mu(R_{1i}) q_2\frac{F_2^2}{m_A^2}  + \sum_{j=2}^{N_2} \mu(R_{2j}) \frac{q_2^2}{q_1} \frac{F_2^2}{m_A^2}\right]\theta_L G^a_{\mu \nu} \tG^{a\mu \nu}. \label{eq:thetaLintermed2}
\end{align}
Notice that we have now written the coupling in terms of a single term that depends on one of the heavy fields, together with a sum over {\em only} the light fields (i.e., the sums omit $i = 1$ and $j = 1$). The first term in brackets in \eqref{eq:thetaLintermed2} is a rational number, while the others are, in general, irrational. However, recall that this is not unexpected: there are additional light fermions in the theory, labeled by $i \in \{2, \ldots, N_1\}$ and $j \in \{2, \ldots, N_2\}$. The quantization condition applies only to a combination of coefficients like \eqref{eq:invariantquantization}, which depends on how the light fermions couple to the axion. By either performing an anomalous field redefinition to eliminate the $\theta_L \Psi \widetilde{\Psi}$ couplings, or computing the one-loop triangle diagram contribution to the $\theta_L G \tG$ amplitude, we find that the light fermion contributions cancel the irrational pieces of the terms in \eqref{eq:thetaLintermed2} that arise from the sum over $i$ and $j$. To compute these contributions, we note that the masses of the $Q_{1i}$ and $Q_{2j}$ fields are proportional to
\begin{align}
\phi_1^\dagger &\sim \exp(-\iu \theta_1) = \exp\left[-\iu \left(q_1 a_H + \frac{q_2 \gcd(q_1,q_2) F_2^2}{m_A^2} \theta_L\right)\right], \nonumber \\
\phi_2^\dagger &\sim \exp(-\iu \theta_2) = \exp\left[-\iu \left(q_2 a_H - \frac{q_1 \gcd(q_1,q_2) F_1^2}{m_A^2} \theta_L\right)\right],
\end{align}
where we have changed to the basis of heavy and light fields. As a result, if we eliminate the $\theta_L$ couplings to the light fermions, we produce new contributions to the $\theta_L G \tG$ coupling,
\begin{align}
\Delta {\cal L}_{\theta_L G \tG} =& - \frac{\gcd(q_1,q_2)}{16\cpi^2} \frac{1}{m_A^2} \left[\sum_{i=2}^{N_1} \mu(R_{1i}) q_2 F_2^2  - \sum_{j=2}^{N_2} \mu(R_{2j}) q_1 F_1^2\right] \theta_L G^a_{\mu \nu} \tG^{a\mu \nu}.  \label{eq:DeltathetaLGtG}
\end{align}
The first of the new terms cancels the middle term in brackets in \eqref{eq:thetaLintermed2}, while the second term combines with the last term in brackets in \eqref{eq:thetaLintermed2} and simplifies:
\begin{align}
{\cal L}_{\theta_L G \tG} + \Delta {\cal L}_{\theta_L G \tG} &= \frac{\gcd(q_1,q_2)}{16\cpi^2} \left[\frac{\mu(R_{21})}{q_1} + \sum_{j=2}^{N_2} \mu(R_{2j}) \left(\frac{q_2^2}{q_1} \frac{F_2^2}{m_A^2} + \frac{q_1 F_1^2}{m_A^2}\right)\right]  \theta_L G^a_{\mu \nu} \tG^{a\mu \nu}\nonumber \\
&= \frac{1}{32\cpi^2} \left[\frac{2 \gcd(q_1,q_2)}{q_1} \sum_{j=1}^{N_2} \mu(R_{2j})\right] \theta_L G^a_{\mu \nu} \tG^{a\mu \nu}.
\end{align}
Now we have finally obtained a manifestly quantized coupling, as we expect for a periodic axion. We can argue that the term in brackets is an integer in precisely the same way that we argued following \eqref{eq:divisibility}, once we make use of \eqref{eq:SUN2U1cancel} and the aforementioned integer quantization of $2\mu(R)$.

\section{Mixing with a Heavier Non-compact Scalar}
\label{sec:non-compact}

In this section, we will study examples in which an axion mixes with a non-compact scalar. As in our previous examples, our purpose is to study the periodicity of the light axion after decoupling the heavy field. In our first example, we consider mixing of the light axion with a radial mode of the same complex field. In the second example, we consider mixing of an ordinary axion with a monodromy axion. For concreteness, we consider an extra-dimensional realization of monodromy in which the two axions are the Wilson loop phases of two different five dimensional gauge fields obtained after compactification on $\mathbb{R}^{3,1} \times S^1$. One has been Higgsed ($H^\mu$), the other remains massless ($A^\mu$), and both couple to the same charged bulk scalar. While there is extensive literature on both the one-loop potential (e.g., \cite{HOSOTANI1983193,ArkaniHamed:2003wu,Feng:2003mk,Delgado:1998qr,Cheng:2002iz,Hatanaka:1998yp,Antoniadis:2001cv}) and axion monodromy (e.g., \cite{Silverstein:2008sg, Mcallister:2008hb, Kaloper:2008fb, Berg:2009tg, Dong:2010in, Kaloper:2011jz, Marchesano:2014mla}), we highlight features of their interplay which have not previously been emphasized in the literature and use them to show our broader conclusions still hold in a more general setting. In both of the examples we consider, if the non-compact field is much heavier than the axion, we find that we can integrate it out to obtain a typical EFT of the light axion. In the case of mixing with a monodromy axion, we find that in the limit where the monodromy potential is subdominant to a periodic potential for a linear combination of the ordinary and monodromy axions, the monodromy is effectively ``realigned'' to the surviving light axion in the EFT, which has a larger decay constant than the original monodromy axion. In every case, we find that deviations of $\theta G \tG$ couplings from their quantized values are, as before, proportional to the mass squared of the axion field.

The case of mixing with a monodromy axion that we discuss is related to an earlier discussion in \cite{Hebecker:2015rya}, in which certain axions obtain masses via fluxes (which makes them monodromy axions) and other axions remain light. That paper emphasized that the light axions can have enhanced field ranges, providing an implementation of alignment \cite{Kim:2004rp} in which the heavy mode is decoupled by fluxes rather than a periodic potential. Our claims are in accord with theirs, but we consider an extended range of possibilities including the scenario when a periodic potential provides a larger mass term than a monodromy potential.

\subsection{Mixing with a Radial Mode}
\label{subsec:radialmixing}

\begin{figure}[!h]
\centering
\includegraphics[width=0.3\textwidth]{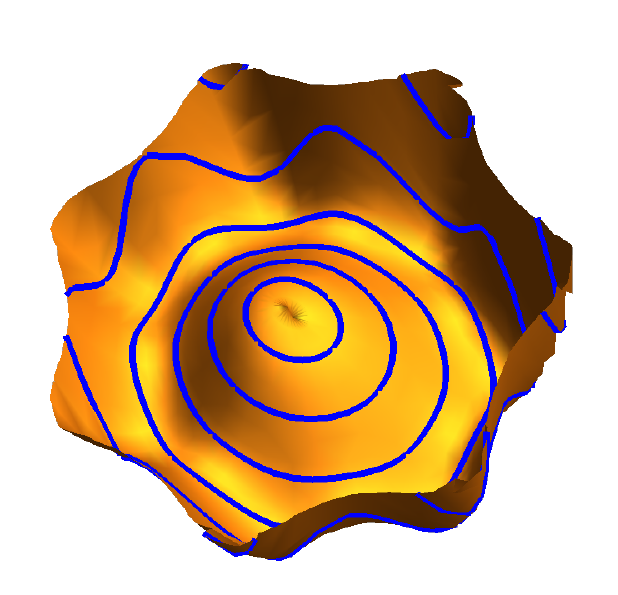}
\caption{An example potential from mixing an axion with a radial higgs mode through a KSVZ-like model, including additional PQ breaking terms. Lines of constant $r$ (defined to be the radial mode of $\phi$) are shown in blue. Both the quark couplings and the PQ-breaking terms generate a potential along the $\theta$ (the phase of $\phi$) direction, but are unable to produce a non-compact valley at the minimum of the potential.}
\label{fig:radialmixing}
\end{figure}

As our first example of mixing with a non-compact scalar, we consider a simple KSVZ UV completion of a single axion \cite{Kim:1979if, Shifman:1979if} and add at least one PQ-breaking term:
\begin{equation}
\Lag = \lambda(|\phi|^2 - v^2)^2 + \left( y \phi Q \widetilde{Q} + \text{h.c.}\right) + \left (\frac{z \phi^N}{\Lambda^{N - 4}} + \text{h.c.}\right).
\end{equation}
The presence of the PQ-breaking term allows, when perturbing around a generic point in field space, for the radial and angular modes of $\phi$ to mix with each other. (The Yukawa term also allows this, after confinement.) An example of this potential for a particular choice of parameters is shown in figure \ref{fig:radialmixing}. Our purpose in studying this theory is to understand whether it can produce a non-compact scalar field after integrating out the radial mode. We find that the answer is no, because even before integrating out the radial mode, we see that there is a nearly-flat, periodic valley at the minimum of the potential. 

The potential, within the UV theory of two real scalar fields, has a periodic valley because of the form of the radial dependence of each of the contributions to the potential. In order for the valley to unwind into a non-compact flat direction, a cross section of the potential at fixed theta would have to oscillate as a function of the radial mode. However, in each of the contributions to the potential in this example, the radial dependence is polynomial. In general, models of this form will generate potentials that are a sum of periodic functions of theta each multiplied by an envelope function that is a polynomial in the radial model. This means that while PQ-breaking terms can generate complicated radial dependence for the precise location of the minimum of the potential, they cannot make the valley unwind into a non-compact direction without fine-tuning the coefficients of the radial polynomials to approximate a periodic function.

\subsection{Mixing with a Monodromy Axion}
\label{subsec:monodromy}

The simple four-dimensional theory that we considered in the previous subsection is not sufficient to allow the valley at the base of the potential to become a non-compact direction. Our discussion suggests that this is more likely to occur in a potential that mixes the radial and angular modes \textit{inside} a periodic function. To generate this type of potential, we will consider the case where a monodromy axion mixes with an ordinary, compact axion. In this model, the nearly-flat valley at the base of the potential is non-compact because of mixing with the non-compact monodromy axion. On the other hand, there is still an underlying periodicity, which is reflected in quantized $a G \tG$ couplings. The mixing between one monodromy axion and a second compact axion can lead to a low energy EFT of a monodromy axion which is lighter and has a larger field range than the original monodromy axion on its own. This type of potential and string theory completions have been discussed previously in \cite{Berg:2009tg}. Instead of a string theory construction, we will instead illustrate the core concepts using a simpler mechanism for generating the potential from dimensional reduction of a higher dimensional QFT.

We consider the simplest toy example that has this effect: 
a five dimensional theory theory of two U(1) gauge fields, where one has been higgsed ($H^\mu$) and the other remains massless ($A^\mu$). The use of massive U(1) gauge fields in higher dimensions to produce monodromy axions in a compactified theory has been discussed previously in \cite{Marchesano:2014mla, Furuuchi:2015foh, Dolan:2017vmn}. After compactifying the fifth dimension on a circle of radius $R$, we obtain axions as the Wilson loop phases $\theta_i \df \oint {\rm d}x^5 G_{5i}$ of the fifth component of each gauge field around the compactified dimension. We can see that the axion is normalized here to have period $2\cpi$ because $G_{5i}$ and $G_{5i} + \frac{1}{R}$ are related by a large gauge transformation. The higgsed field alone will generate a monodromy potential for $\theta_{H}$. To generate a periodic potential that mixes $\theta_H$ and $\theta_A$, we couple both gauge fields to the same form of matter.  For simplicity, we take this matter to be a massless scalar and take the 5d spin-1 field to have a simple St{\"u}ckelberg mass term, but these choices do not qualitatively change our results. (In particular, our qualitative conclusions should carry over to the other shapes of monodromy potentials that are known to arise in string models, e.g., \cite{Silverstein:2008sg, Dong:2010in, McAllister:2014mpa}.) The action in this theory takes the form
\begin{equation}
S = \int {\rm d}^5 x \left(-\frac{1}{4g_{5H}^{2}} H_{MN}(x) H^{MN}(x) - \frac{m^2}{2g_{5H}^2} \mathcal{H}_\mu \mathcal{H}^\mu -\frac{1}{4g_{5A}^2} A_{MN}(x)A^{MN}(x) + D_M \chi^\dag(x) D^M \chi(x)\right)
\end{equation}
where the covariant derivative is
\begin{equation}
D_M \chi(x) \df \p_M \chi(x) - \iu q_A A_M(x) \chi(x) - \iu q_{H} H_M(x) \chi(x)
\end{equation}
and following \cite{Furuuchi:2015foh} we have defined
\begin{equation}
\mathcal{H}_M(x) \df H_M(x) - \iu e^{\iu \theta(x)} \p_M \E^{-\iu \theta(x)},
\end{equation}
where the St{\"u}ckelberg field $\theta(x)$ is a periodic scalar. Since $\theta$ is an angular variable it can have nontrivial winding around the extra dimension, $\frac{w x^5}{R}$ for integer $w$, which is responsible for the monodromy after compactification.

The potential obtained after compactification contains two distinct contributions. At tree level, we only see the monodromy potential of the higgsed gauge field from the mass terms \cite{Furuuchi:2015foh}
\begin{equation}
\Lag_4 \supset -V_\text{mon}(\theta_H) \df -\frac{m^2}{2 g_{4H}^2 R^2} \left(\frac{\theta_H}{2\cpi} - w\right)^2 = - \frac{1}{2} m^2 F_H^2 \left(\theta_H - 2\cpi w\right)^2,
\end{equation}
where we have defined the 4d gauge couplings $g_{4i} = g_{5i}/\sqrt{2\cpi R}$ as well as the decay constants of the 4d axion fields, $F_i = 1/(2\cpi g_{4i} R)$. Since the kinetic terms are $\frac{1}{2} F_i^2 (\partial \theta_i)^2$, we see that $m$ is the canonically normalized mass of $\theta_H$. As is typical with monodromy, for the Lagrangian to remain invariant under a shift by the axion period, we must also shift $w$. On any given branch of fixed $w$, the potential is effectively not periodic, so $\theta_H$ behaves as a non-compact scalar. In addition to the tree level potential, both gauge fields get one-loop potentials from their couplings to matter. Since they are coupled to the same form of matter, the one loop potential will be a periodic potential that mixes the monodromy axion with the ordinary axion. In particular, the potential generated by integrating out the mass terms for the tower of scalar Fourier modes
\begin{equation}
\chi^{\dag(n)} \frac{1}{R^2}\left(n - q_A \frac{\theta_A}{2\cpi} - q_H \frac{\theta_H}{2\cpi}\right)^2 \chi^{(n)}
\end{equation}
will simply be a sum of cosines in the case where $\chi$ is massless (e.g., \cite{ArkaniHamed:2003wu}),
\begin{equation}
V_{\rm per}(\theta_A, \theta_H) = -\frac{3}{64 \cpi^6 R^4}\sum_{n = 1}^\infty \frac{\cos(n q_A \theta_A + n q_H \theta_H)}{n^5}.
\end{equation}
In the case where $\chi$ is massive the exact form the potential is more complicated \cite{HOSOTANI1983193,Delgado:1998qr,Feng:2003mk}, but will still be periodic and produce qualitatively the same effect. 

\begin{figure}[!h]
\centering
\includegraphics[width=0.4\textwidth]{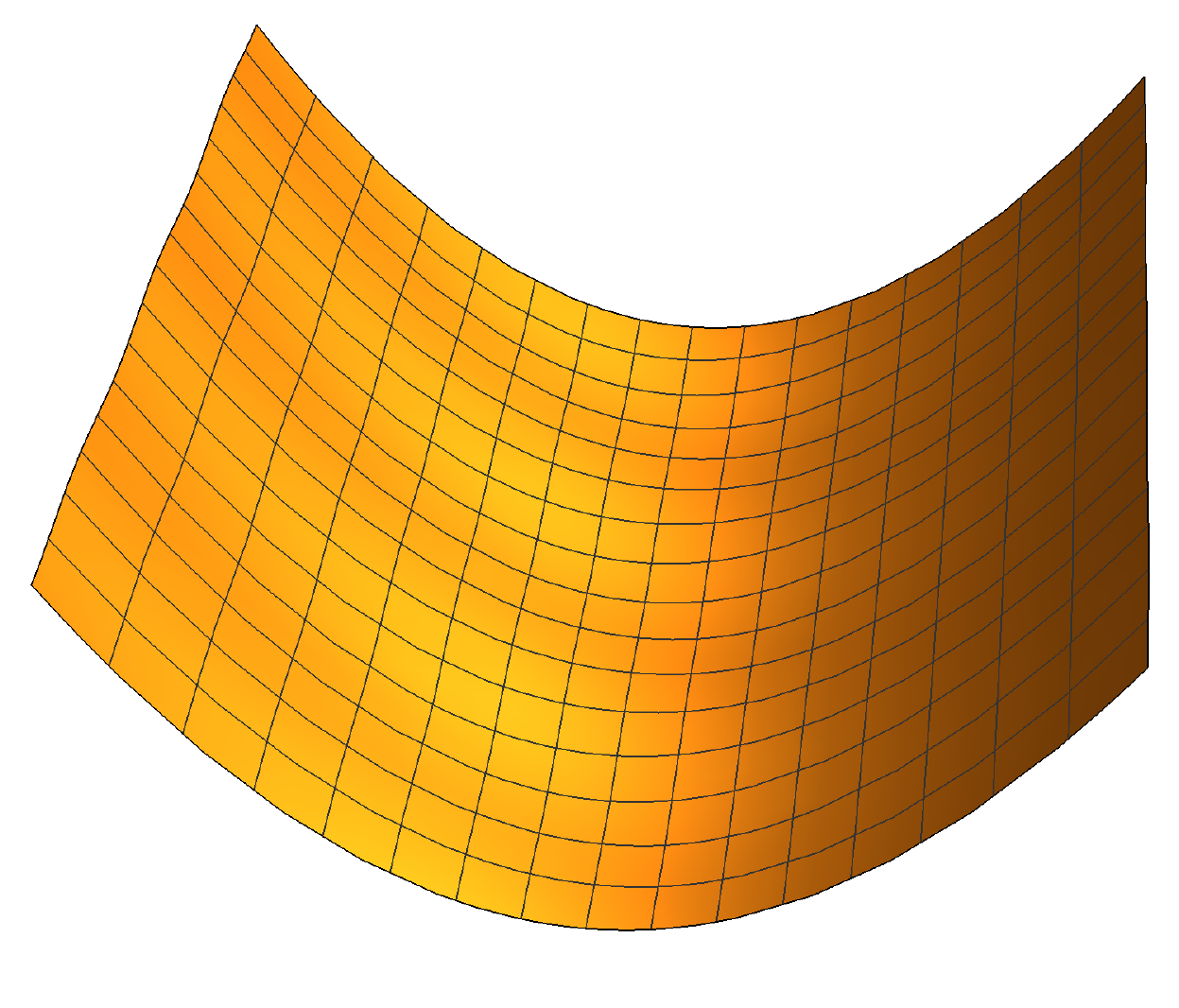}\quad
\includegraphics[width=0.4\textwidth]{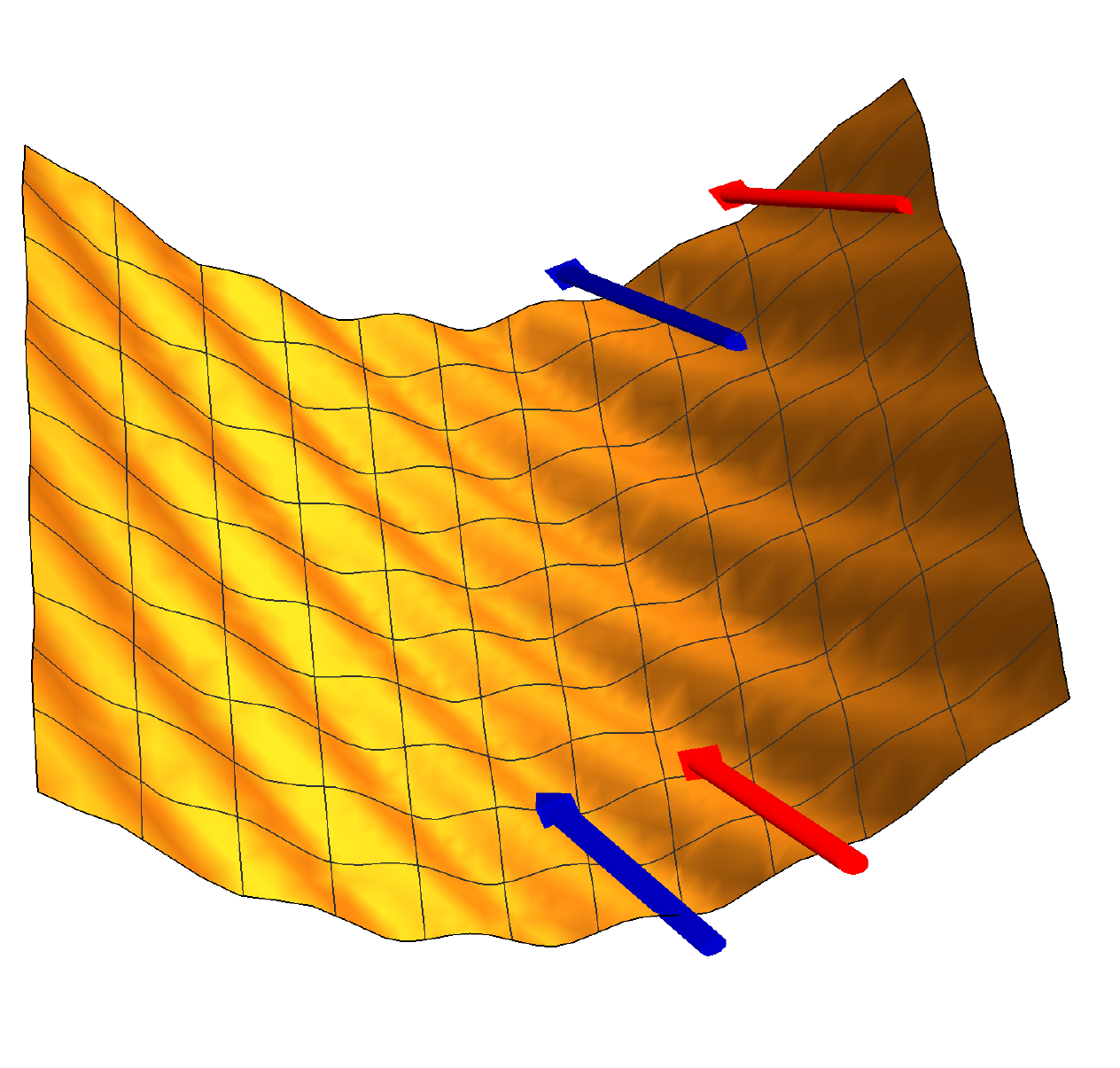}
\caption{The two-axion potential in the case where a monodromy axion $\theta_H$ (horizontal axis) mixes with a compact axion $\theta_A$ (vertical axis). The front and back edges of the surface correspond to $\theta_A = 0$ and $\theta_A = 2\cpi$ and are identified. In the left plot, we have chosen parameters so that $\partial_H^2 V_\text{mon} > |\partial_H^2 V_\text{per}|$. In this case, the monodromy axion is heavy and can be integrated out; the light axion $\theta_A$ has a smaller cosine potential along the periodic valley at the base of the monodromy axion's potential. In the right plot, the opposite limit $|\partial_H^2 V_\text{per}| > \partial_H^2 V_\text{mon}$ is taken. In this case, the cosine potential is large enough to produce a series of ridges. The light axion mode is neither $\theta_A$ nor $\theta_H$, but the mode that traverses the valley in between ridges, along which $\theta_A = -q_H \theta_H/q_A\, (\text{mod}\, 2\cpi)$. The colored arrows show the path of the minimum down the potential, where arrows of a particular color should be identified together.}
\label{fig:twoaxionmonodromy}
\end{figure} 

Although $V_\text{per}$ is a one-loop effect and $V_\text{mon}$ is a tree-level effect in this model, it need not be the case that the monodromy potential dominates. This is because $V_\text{mon}$ originates from spontaneous breaking of the discrete shift symmetry, which is preserved by $V_\text{per}$, so it is of parametrically different (potentially much smaller) size. It is interesting to consider two different limits, one in which $\partial_H^2 V_\text{mon} > |\partial_H^2 V_\text{per}|$ throughout the field space, and one with the opposite inequality. (Here $\partial_H$ denotes $\partial/\partial \theta_H$.) These are depicted in the left- and right-hand panels of Fig.~\ref{fig:twoaxionmonodromy}, respectively. The left panel shows the case where the monodromy potential dominates over the periodic potential. The periodic potential creates a small perturbation, but there is no obstruction to any nonzero value of $\theta_H$ rolling down the potential toward $\theta_H = 0$. The right panel shows the more interesting case, in which $|\partial_H^2 V_\text{per}| > \partial_H^2 V_\text{mon}$. This creates a series of ridges in the potential; it is conceivable that the field could be localized (for instance, during inflation) in a valley between ridges far up the potential, and will evolve toward the minimum by following the winding path down the valley rather than moving directly in the $\theta_H$ direction.

The phenomenon exhibited in the case with a ridged potential might be thought of as ``monodromy realignment.'' In the effective theory containing both $\theta_H$ and $\theta_A$, it is $\theta_H$ that carries the monodromy. This is because the St{\"u}ckelberg field that produced the monodromy shifted only under shifts of $H_\mu$. Nonetheless, the low-energy effective theory is that of a monodromy axion that is a nontrivial linear combination of $\theta_H$ and $\theta_A$. One way to see this is by noting that we could first integrate out the linear combination of fields that obtains a mass from $V_{\rm per}$. As in \S\ref{subsec:latticebasis}, we could choose an alternative lattice basis $(\theta_1, \theta_2)$ in which this field is $\theta_1$. Specifically, we can find integers $r_A, r_H$ such that
\begin{equation}
\begin{pmatrix} \theta_1 \\ \theta_2 \end{pmatrix} = \begin{pmatrix} p_A & p_H \\ r_A & r_H \end{pmatrix} \begin{pmatrix} \theta_A \\ \theta_H \end{pmatrix},
\end{equation}
where
\begin{equation}
p_A \df \frac{q_A}{\gcd(q_A,q_H)}, \quad p_H \df \frac{q_H}{\gcd(q_A,q_H)}, \quad \text{and} \quad p_A r_H - p_H r_A = 1.
\end{equation}
In this basis, the potential is (choosing the branch where $w = 0$)
\begin{equation}
V(\theta_1, \theta_2) = V_{\rm per}(\theta_1) + \frac{1}{2} m^2 F_H^2 \left(p_A \theta_2 - r_A \theta_1\right)^2.
\end{equation}
The effective theory along the valley in the potential is obtained by taking $\theta_1 = 0$ (or a $2\cpi$ shift thereof), so that we can integrate it out to obtain an effective theory of the light field $\theta_2$,
\begin{equation}
\frac{1}{2} F_2^2 \partial_\mu \theta_2 \partial^\mu \theta_2 - \frac{1}{2} m^2 p_A^2 F_H^2 \theta_2^2 + (\text{terms proportional to }\Box \theta_2),
\end{equation}
where, using \eqref{eq:latticebasistransform}, the kinetic term of $\theta_2$ is proportional to
\begin{equation}
F_2^2 = p_A^2 F_H^2 + p_H^2 F_A^2.
\end{equation}
From this we can read off the canonically normalized mass of the light field,
\begin{equation}
m_2^2 = m^2 \frac{p_A^2 F_H^2}{p_A^2 F_H^2 + p_H^2 F_A^2}.
\end{equation}
The nonperiodic potential for $\theta_2$ indicates that, in the low-energy effective theory, it is a monodromy axion; we say that the monodromy has {\em realigned} from $\theta_H$ to $\theta_2 = r_A \theta_A + r_H \theta_H$. Monodromy realignment has both {\em increased} the effective decay constant and, correspondingly, {\em decreased} the mass of the monodromy axion. Both of these features are intuitively apparent from the winding valley in Fig.~\ref{fig:twoaxionmonodromy}.

We could also ask if couplings to external gauge fields are quantized the way that we expect them to be. To study this we consider adding Chern-Simons terms to the theory,
\begin{equation}
\Lag_{CS} = \frac{c_A}{16 \cpi^2} \epsilon^{MNPQR} A_M \Tr[G_{NP}G_{QR}] + \frac{c_H}{16 \cpi^2} \epsilon^{MNPQR} H_M \Tr[G_{NP}G_{QR}],
\end{equation}
where $G$ is an arbitrary gauge field (which could be one of the two already in the theory). Gauge invariance requires that the coefficients $c_i$ be integers. After dimensionally reducing, these Chern-Simons terms will contain $\theta_i G_\mu \tG^\mu$ couplings of the axions to the four dimensional gauge fields with quantized couplings: gauge invariance required us to start with $c_i$ quantized, and dimensionally reducing won't change that. Just as in earlier sections, the change of lattice basis from $(\theta_A, \theta_H)$ to $(\theta_1, \theta_2)$ does not change the quantization of the $\theta G \tG$ couplings. However, even though we chose our $(\theta_A, \theta_H)$ basis to have diagonal kinetic terms (which need not be true, in general), the kinetic terms in the $(\theta_1, \theta_2)$ basis are generally not diagonal. As in \S\ref{subsec:periodicityEFT}, when we integrate out $\theta_1$, we will generally obtain terms $\propto (\Box \theta_2) G \tG$ in the low-energy EFT. When we consider the mass that $\theta_2$ obtains from $V_{\rm mod}$, these will appear as effectively non-quantized couplings. Just as in our earlier discussion, these contributions are all proportional to the mass parameter $m_2^2$ of the light axion.

\section{Non-compact Symmetries Should Not Emerge in the IR}
\label{sec:gaugetheorycompare}

We can summarize our results by saying that if we start with a theory of several axions and, in one way or another, decouple some linear combinations of them while leaving others massless, the massless fields will still be axions, i.e., their field space will be compact and their couplings will be quantized accordingly. In cases where we found non-quantized couplings of a light axion field, we found that the field also obtained a mass, and the deviation of the axion's couplings from their quantized values were proportional to the mass squared of the axion. As we noted in the introduction, this has the same flavor as a well-known fact about gauge theory: if we begin with a compact gauge group and then Higgs it, the surviving infrared gauge group will be compact (and hence will have quantized charges). Such a result is known to hold in many different contexts with compact gauge groups in the UV, in cases where we decouple gauge fields via confinement, via Chern-Simons mass terms in (2+1)d gauge theory, or even when we alter the gauge group entirely in the infrared, as in Seiberg duality. It is also known to be robust against kinetic mixing  \cite{Holdom:1985ag, Shiu:2013wxa}.

Our observations about axions and the corresponding observations about gauge fields are linked in more than a vague qualitative manner. In the case of (2+1)d theories, they are identical, because a massless axion field $\theta$ in (2+1)d is Hodge dual to a gauge field $A_\mu$ defined by ${\rm d}A = 2\cpi e F_\theta \star {\rm d}\theta$, where $2\cpi F_\theta$ is the distance in field space around the $\theta$ circle and $e$ is the gauge field coupling. The scenario discussed in \S\ref{subsec:spin1diag}, where $\theta$ is eaten to provide a St{\"u}ckelberg mass to another gauge field $B$, maps to precisely the case where the gauge field $A$ dual to $\theta$ obtains a mass through a mixed Chern-Simons term $B \wedge {\rm d}A$. The low-energy theory contains a massless gauge field for a compact gauge group with finite coupling, which is dual to a compact axion field.

One reason to expect that a theory with a compact gauge group in the UV flows to a theory with a compact gauge group in the IR is that any effective field theory that contains a non-compact gauge group, such as $\mathbb{R}$, is believed to be inconsistent when coupled to gravity. In such theories, one can generally construct black holes of irrational charge \cite{Banks:2010zn}, which violate entropy bounds that are believed to be true in all theories of quantum gravity \cite{Bousso:1999xy}. If it were possible to construct UV theories with compact gauge groups that flow to IR theories with non-compact gauge groups, the UV theory would lie in the Swampland \cite{Vafa:2005ui}. This would be an interesting new Swampland constraint, but we are unaware of any examples that realize such RG flows.

One possible reason why such RG flows do not exist in general is that they lead to IR theories with a continuum of operators that did not exist in the UV. In theories with a compact gauge group that has an associated $p$-form gauge field $A_p$, Wilson line or surface operators of the form $\exp(\iu q \int_\Sigma A_p)$, where $\Sigma$ is a $p$-dimensional submanifold of spacetime, are defined for discrete choices of charge $q \in \bbZ$. If the gauge group is $\mathbb{R}$, then there is a {\em continuum} of well-defined operators with arbitrary $q$. A similar statement holds for axions: if $\theta$ is a $2\cpi$-periodic boson, then $\theta$ itself is not a well-defined operator, but $\exp(\iu q \theta)$ for $q \in \bbZ$ is a sensible local operator. On the other hand, in the non-compact limit, there is no obstruction to constructing such operators for arbitrary $q \in \mathbb{R}$. This suggests a possible general argument against the emergence of either non-compact gauge groups or non-compact bosons from theories with compact gauge groups and axions in the UV: this would be an RG flow from a UV theory with a discrete operator spectrum to an IR theory with a continuous operator spectrum. It seems plausible that such RG flows are forbidden in sensible theories.

In this paper, we will not go further in attempting to make these suggestions rigorous, but we believe that they point toward a deeper understanding of why our results hold. The properties that arise in many different effective field theories of axions are very closely akin to properties arising in gauge theories, and are likely to be enforced by very general principles of quantum field theory.

\section{Conclusions}
\label{sec:conclusions}

Periodicity imposes strong constraints on the axion couplings and field ranges, even in cases where axions mix with other axions or a non-compact scalar. Given our results, it appears the options for generating significantly different axion couplings or field ranges than naively expected are: generating a large integer in the effective theory of a single light axion, as in the clockwork scenario \cite{Kim:2004rp, Dvali:2007hz, Choi:2014rja, Hebecker:2015rya, Choi:2015fiu, Kaplan:2015fuy}; building an effective theory that intrinsically involves multiple axions (e.g., kinetically mixing the axion of interest with an even lighter one); or relaxing these constraints through effects proportional to the mass of the light axion (e.g., realignment of monodromy). While the clockwork scenario has been explored extensively, further studying kinetic mixing with a lighter axion and realignment of monodromy could have potentially interesting phenomenological prospects.

\section*{Acknowledgments}

We thank JiJi Fan for a useful comment on the draft. KF is supported by the National Science Foundation Graduate Research Fellowship Program under Grant No.~DGE1745303. MR is supported in part by the DOE Grant DE-SC0013607.

\bibliography{ref}
\bibliographystyle{utphys}

\end{document}